\documentclass[preprint, authoryear, 3p, 12pt]{elsarticle}

 \pdfoutput=1 

\usepackage{amsmath}
\usepackage{amsfonts}
\usepackage{amssymb}
\usepackage{breqn}
\usepackage{subfigure}
\usepackage{xcolor}
\usepackage{cleveref}
\usepackage{amssymb,amsmath,bm,bbm}

\begin{document}

\begin{frontmatter}

\title{Discrete defect plasticity and implications for
  dissipation\footnote{Based on a lecture presented at the 11th 
European Solid Mechanics Conference, 4-8 July 2022, Galway, Ireland}}

\author{Alan Needleman}

\address{Department of Materials Science \& Engineering,
 Texas A\&M University,  College Station, TX 77843 USA}

\begin{abstract}
Inelastic deformation of solids is almost always, if not always,
associated with the evolution of discrete defects such as
dislocations, vacancies, twins and shear transformation zones. The
focus here is on  discrete defects that can be
modeled appropriately by continuum mechanics, 
but where the discreteness of the carriers of plastic
deformation plays a significant role. The formulations  are
restricted to small deformation kinematics and the defects considered,
dislocations and discrete shear transformation zones (STZs),  
are described by their linear elastic fields. In discrete defect
plasticity both the stress-strain response and the
partitioning between defect energy storage and defect  dissipation are
outcomes of an initial/boundary value problem solution. 
For such defects an explicit expression for the dissipation rate  is
presented and, because there is 
a length scale associated with the discrete defects, the
stress-strain response and  the evolution of the dissipation rate 
can be size dependent.  Discrete dislocation plasticity modeling
results are reviewed that 
illustrate  the implications of  defect dissipation evolution for
friction, fracture, fatigue 
crack growth and thermal softening.
 Examples are also given of the consequences  in constrained shear of
three modes of the evolution of 
discrete defects for the size  dependence of the stress-strain
response and the associated  dissipation.  
For a
purely mechanical formulation the Clausius-Duhem inequality
specializes to the requirement that the dissipation rate  is
non-negative. Requiring a non-negative dissipation rate for all points
of a body and for all time
imposes restrictions on kinetic relations for  
the evolution of  discrete defects.  In  statistical
mechanics,  the Clausius-Duhem inequality can be violated  for 
sufficiently small  regions for a sufficiently  short  time.  Explicit
kinetic relations for discrete 
dislocation plasticity dissipation and for 
discrete STZ plasticity dissipation identify conditions that can lead
to a negative dissipation rate.  In
continuum mechanics,  at least in some circumstances, satisfaction of
the Clausius-Duhem inequality can be regarded as a stability
requirement. Nevertheless, simple one-dimensional  continuum
calculations  illustrate that there can be a 
negative dissipation rate over a small region  and
for a short time period with overall stability maintained. 
Implications for discrete defect plasticity modeling are briefly discussed. 
\end{abstract}


\end{frontmatter}

\section{Introduction}

Inelastic deformation of solids generally takes place by the evolution in space
and time of a collection of discrete defects such as dislocations,
vacancies, phase transformations, twins, and shear transformation
zones (STZs).  A variety of
deformation and failure processes are governed by the evolution of
collections of such defects, in many circumstances each of 
which can be modeled appropriately by continuum mechanics,
but where the discreteness of the carriers of plastic
deformation plays a significant role. Discrete defect effects can be
key for modeling structure/component performance and reliability for
both micro-scale and macro-scale components and devices.  In
particular, discrete defects  dissipate
energy and store energy, and the balance between 
dissipation and energy storage plays a key role in a variety of
phenomena of technological significance including, for example,
thermal softening, friction, fracture and fatigue.  

As for any continuum formulation, in addition to  satisfying basic 
principles, i.e.  the kinematics of deformation and the conservation
of mass, momentum and energy, constitutive relations (in this context
termed kinetic relations) are needed for modeling 
the discrete defects. In discrete defect plasticity, the stress-strain
response is an outcome of an initial/boundary value problem solution
(not an input as in conventional continuum plasticity) as is the
partitioning between defect energy storage and defect  dissipation. 
Furthermore, there
is a length scale associated with discrete defects, for example, the
Burgers vector for a dislocation or  size associated with
a shear transformation zone (STZ), so that both the stress-strain response
and the evolution of the energy partitioning can be size dependent. 

A  proper continuum mechanics
formulation is also expected to satisfy a  statement (or perhaps more
accurately one of the statements\footnote{``There 
 have been nearly as many formulations of the second law 
as there have been discussions of it.'' P.W. Bridgman})  of the second
law of thermodynamics. 
 The statement of the second law in terms of  satisfying the 
Clausius-Duhem inequality for all points of a body for all
time was introduced as a postulate into continuum mechanics by
\cite{CN64} and provides a formalism for developing constitutive
relations consistent with that inequality for all time.
In discrete defect plasticity, satisfaction of the
Coleman-Noll postulate, i.e. requiring the
dissipation rate to be non-negative for all material regions and for
all time,  places a restriction on the kinetic relations 
for defect evolution. 
 
  Satisfaction of the Coleman-Noll postulate \citep{CN64} is
 well-established for macro-scale continuum modeling.
  In statistical mechanics this requirement emerges for large
 systems and long periods of time, \cite{Evans02,Jarz10}.  However,  theoretical, 
 computational  and experimental studies have
 shown that,  when  the 
process under consideration involves  a sufficiently small  
 region  for a  sufficiently  short time,   satisfying the
 Clausius-Duhem inequality  has a probability less than one,
 e.g. \cite{Evans93,Ayton01,Evans02,Jarz10, Wang10, Ostoja20}.
  This may be relevant for 
 discrete defect plasticity since the change in defect state may only
 involve a small number of discrete entities in a sufficiently small
 material region and
  may occur over a sufficiently short
 time.  Key issues, of course, are  what  is ``sufficiently small''
and  what  is ``sufficiently short.'' 

 We begin with a brief presentation of the Clausius-Duhem inequality
and illustrate 
the implications of requiring satisfaction of the Coleman-Noll
postulate \citep{CN64} for constitutive relations in conventional continuum
mechanics.  A  
continuum mechanics framework is then outlined for solving problems
with plastic deformation arising  from the evolution of 
discrete defects. The continuum mechanics framework is restricted to
small deformations (i.e. geometry changes neglected) and attention is
focused on  defects, such as dislocations and shear transformation
zones (STZs), that can be described in terms of their  linear elastic
fields;  $1/r$ singular strain/stress fields for dislocations and
\cite{Eshelby57,Eshelby59} inclusion fields for STZs. For
such defects, dissipation only occurs on surfaces with 
displacement/displacement  rate  jumps.  

Explicit expressions for the
dissipation rate for discrete dislocation
plasticity and for discrete STZ plasticity are given. Particular attention is
directed to restrictions imposed on discrete 
defect kinetic relations by requiring the dissipation rate to be
non-negative. 
In addition,  results are presented of discrete dislocation
plasticity analyses that illustrate  the implications of
the evolution of  defect dissipation for  friction, fatigue crack
growth and the development of the Bauschinger effect  and thermal
softening in plane strain tension.   These results  illustrate what can
be predicted by discrete dislocation modeling, with the dissipation
rate being non-negative for all time, for the evolution of
dissipation,  for its size dependence
and for the resulting  effect on material response.

 For STZ plasticity, the requirement of a non-negative dissipation
 rate for all time restricts the  
transformation strain magnitude  to be smaller than
indicated by atomistic analyses and experiment,
e.g. \cite{Argon,Dasgupta,Albert,Hufn16}.
Possible reasons for this discrepancy include:  (i) nonlinear
elastic/non-elastic material behavior;   (ii) identifying the
\cite{Eshelby57,Eshelby59}  inclusion transformation stress with the yield strength
and not accounting for possible local stress concentrations; (iii)
neglecting the role of entropy, particularly configurational entropy;
and (iv) this could 
be a circumstance where the Clausius-Duhem
inequality can be violated for a small number of discrete entities and
a short time. 

Discrete  
STZ  plasticity predictions  are  presented for the stress-strain
response and for the 
spatial stress, strain and dissipation distributions  in plane strain
tension with the formulation satisfying the requirement of a
non-negative dissipation rate  
for all STZs for all time. 

Examples are  given of consequences for the size dependence of the
stress-strain response and for the size dependence of plastic
dissipation of three modes of  evolution
of discrete defects: (i) relatively few defects that
move relatively long 
distances through the material;  (ii) nucleation of
defects from fixed nucleation sites with the nucleated defects moving
only short distances or not all; and (iii) nucleation of  defects
that do not move but that induce a percolation-like increase in the
number of nucleation
sites. Each of these leads to a different prediction for the size
dependence of the stress-strain response and of the 
associated dissipation.

The more general question of the consequences of violating 
Clausius-Duhem inequality  in a continuum formulation  (for a purely
mechanical framework  violating 
the requirement of a non-negative 
dissipation rate) for a short time
is explored  in a simple one-dimensional context and possible
implications for discrete defect 
plasticity modeling are briefly discussed.

 \section{The Clausius-Duhem inequality and dissipation}

Attention is restricted to small deformation kinematics and
quasi-static processes (kinetic energy negligible).  Using
Cartesian tensor notation, the strain tensor components,
$\epsilon_{ij}$, are given by 
\begin{equation}
\epsilon_{ij} = \frac{1}{2} \left ( u_{i,j}+u_{j,i} \right )
\label{cd3}
\end{equation}
with $u_i$  the components of the displacement vector,
and $(\_)_{,i}$ denotes $\partial (\ )/\partial x_i$. Also, the stress
tensor components, $\sigma_{ij}$  satisfy
\begin{equation}
\sigma_{ij,j}=0
\label{cd5}
\end{equation}
so that
\begin{equation}
\int_V \sigma_{ij} {\dot\epsilon}_{ij} dV= \oint_S \sigma_{ij} n_j
\dot{u}_i dS
\label{cd7}
\end{equation}
where $V$ is the volume of the body, $S$ is the surface of $V$, $n_j$
are the components of the 
unit normal to $S$ and $(\dot{ \ })=\partial{(\ )}/{\partial t}$.

The only
non-mechanical fields considered are the temperature $\Theta$, the entropy
$s$, and the heat flux vector ${\bf q}$. In this context, the
Clausius-Duhem inequality for a body with volume $V$ then can be
written as \citep{CN64}  
\begin{equation}
\dot{W}-\dot{\Phi}+\int_V \Theta \dot{s} dV -\int_V \left ( \frac{q_i}
  {\Theta}  \right ) \Theta_{,i}  dV \ge 0
\label{cd1}
\end{equation}
where $\dot{W}$ is the rate of working and $\dot{\Phi}$ is the time
derivative of the internal energy, with
\begin{equation}  
\dot{W}=\oint_S \sigma_{ij} n_j
\dot{u}_i dS=\int_V \sigma_{ij} {\dot\epsilon}_{ij} dV = \int_V \dot{w} dV 
  \quad , \quad \dot{\Phi}=\int_V \dot{\phi} dV
\label{cd2}
\end{equation}
In the context of a small deformation framework $\dot{w}$ and
$\dot{\phi}$ are taken to be per unit volume (rather than per unit mass).

If the Clausius-Duhem inequality, Eq.~(\ref{cd1}), is presumed to hold for
any admissible deformation and thermal history, and 
for any sub-volume, then at each point of  $V$ 
\begin{equation}
\sigma_{ij} {\dot\epsilon}_{ij} -\dot{\phi} + \Theta \dot{s} -\left (
\frac{q_i}{\Theta} \right ) \Theta_{,i}  \ge 0
\label{cd4}
\end{equation}
 Note that if Eq.(\ref{cd4}) holds at   then at each point of
  $V$, then Eq.(\ref{cd1}) holds. However, Eq.(\ref{cd1}) can hold
  even if  Eq.(\ref{cd4}) does not hold at each point of
  $V$.

  One example of the implications of requiring satisfaction of the
Clausius-Duhem inequality, Eq.~(\ref{cd4}), for all points of the
body  and for all time, i.e. satisfying the Coleman-Noll
postulate \citep{CN64}, is illustrated by
taking $\phi(\epsilon_{ij},s)$ so that
\begin{equation}
\dot{\phi}=\frac{\partial \phi}{\partial \epsilon_{ij}}
\dot{\epsilon}_{ij} + \frac{\partial \phi}{\partial s} \dot{s}
\label{cd8}
\end{equation}

Substituting Eq.~(\ref{cd8}) into Eq.~(\ref{cd4}) gives
\begin{equation}
\left [ \sigma_{ij} -\frac{\partial \phi}{\partial \epsilon_{ij}}
  \right ] {\dot\epsilon}_{ij} + \left [\Theta- \frac{\partial
    \phi}{\partial s} \right ] \dot{s} -\left (
\frac{q_i}{\Theta} \right ) \Theta_{,i}  \ge 0
\label{cd10}
\end{equation}

Because Eq.~(\ref{cd10}) must hold for any admissible history, it
must hold if any two of ${\dot\epsilon}_{ij}$, $\dot{s}$, $\Theta_{,i}$
vanish. Hence, Eq.~(\ref{cd10}) requires
\begin{equation}
\sigma_{ij} = \frac{\partial \phi}{\partial \epsilon_{ij} }\quad ,
  \quad \Theta= \frac{\partial  \phi}{\partial s} \quad , \quad 
-\left (\frac{q_i}{\Theta} \right ) \Theta_{,i}  \ge 0
\label{cd11}
\end{equation}

With the focus  on a purely mechanical framework so that 
 entropy and heat flux/temperature gradient effects are not
  included in the formulation,  Eq.~(\ref{cd1})
simplifies to 
\begin{equation}
\dot{\cal D}=\dot{W} -\dot{\Phi} \ge 0
\label{cd12x}
\end{equation}
and Eq.~(\ref{cd4}) becomes
\begin{equation}
\dot{\xi}=\sigma_{ij} {\dot\epsilon}_{ij} -\dot{\phi} \ge 0
\label{cd12}
\end{equation}
where $\dot{\cal D}$ is the dissipation rate in the body and
$\dot{\xi}$ is the dissipation per unit volume. 

We consider a class of elastic-plastic or an elastic-viscoplastic
solids for which 
\begin{equation}
\dot{\epsilon}_{ij}=\dot{\epsilon}^e_{ij} + \dot{\epsilon}^p_{ij}
\label{cd13}
\end{equation}
and 
\begin{equation}
\dot{\epsilon}^p_{ij}=\dot{\Lambda} q_{ij} 
\label{cd14}
\end{equation}

Assuming that plastic deformation makes no contribution to the
internal energy so that $\dot{\phi} =\sigma_{ij}
\dot{\epsilon}^e_{ij}$, the local Clausius-Duhem inequality, Eq.~(\ref{cd12}),
then requires  
\begin{equation}
\dot{\xi}=\sigma_{ij} \dot{\epsilon}^e_{ij} + \sigma_{ij}
\dot{\epsilon}^p_{ij}  -\dot{\phi} = \sigma_{ij}
\dot{\epsilon}^p_{ij} =
\dot{\Lambda} \sigma_{ij} q_{ij}  \ge 0
\label{cd15}
\end{equation}
and the Clausius-Duhem inequality is satisfied for all deformation
histories and all time if $\dot{\Lambda} \ge 0$ and $\sigma_{ij}
q_{ij} \ge 0$. 

For example, if
\begin{equation}
q_{ij}=\sigma_{ij}^\prime+\alpha \sigma_{kk} \delta_{ij}
\label{cd15x}
\end{equation}
so that
\begin{equation}
\sigma_{ij} q_{ij}= \left [ \sigma_{ij}^\prime \sigma_{ij}^\prime +\alpha
  \sigma^2_{kk} \right ]
\label{cd15y}
\end{equation}
where $\delta_{ij}$ is the Kronecker delta and $\sigma_{ij}^\prime =
\sigma_{ij} -1/3 \sigma_{kk} \delta_{ij}$, then with $\dot \Lambda \ge 0$,
the Clausius-Duhem inequality is satisfied for all stress states if
$\alpha \ge 0$. 

The statement that the Clausius-Duhem inequality, Eq.~(\ref{cd4}),
must be satisfied for all points of a body and for all deformation and
thermal histories is basically taken as a postulate by \cite{CN64} so
the question arises 
as to what are the implications of it not being satisfied. It has been
shown that satisfaction of the Clausius-Duhem inequality
implies stability\footnote{``Stability is a word with an
unstable definition.'' R. Bellman} 
at least in some cases, for example, 
Lyapunov stability for systems with time evolution governed by
ordinary differential equations,  \cite{Cole67}, 
or continuous dependence of thermodynamic processes on the
initial state and supply terms terms for thermoelastic materials,
\cite{Dafer79}.  A possibility is that more generally  violation of the
Clausius-Duhem inequality (in a purely mechanical
formulation having a negative dissipation rate)  will lead to some
kind of instability.

\section{Discrete defect plasticity}

At  a  size scale where the discreteness of 
defects plays a role, plastic deformation is inevitably
nonuniform.  In a wide variety of circumstances, the
evolution of collections of such defects can be modeled appropriately
within a continuum mechanics framework. Within such a framework, 
the stress strain response and the evolution of the material
microstructure are direct outcomes of an initial/boundary value
problem solution.  A possible size effect of the stress-strain
response and of the energy partitioning, as well as their dependence on stress
state, loading rate, deformation history, etc. is also an outcome.  

The aims of discrete defect plasticity analyses are: (i) to provide predictions 
of the mechanical response of materials that are
difficult, if not impossible, to obtain from smaller scale or larger
scale frameworks; and/or  (ii) to provide guidance
for developing continuum plasticity constitutive relations. 

 Although the general formulation applies to a range of discrete
defects,  attention  here is focused 
dislocations and shear 
transformation zones (STZs). Both are modeled in terms of their linear
elastic fields but have quite different behaviors. Dislocations are
mobile and give rise to 
long range stresses. On the other hand, the positions of STZs are
fixed and their 
associated stress fields have a shorter range than those of
dislocations. 

\subsection{Modeling framework}
\label{sec:model}

In addition to consideration being restricted to small
deformation kinematics  and quasi-static deformation histories,
in the specific situations considered, the stress and strain are taken to be
related by linear isotropic elasticity, so that
\begin{equation}
\sigma_{ij} = L_{ijkl} \epsilon_{kl}=\frac{E}{(1+\nu)} \left [
  \epsilon_{ij} + \frac{\nu}{1-2\nu} \epsilon_{kk} \delta_{ij} \right ]
\label{eqm3}
\end{equation}
with $L_{ijkl}=L_{klij}$ the components of the tensor of elastic moduli, $E$
Young's modulus, $\nu$ Poisson's ratio. Also, for a purely mechanical
formulation, the elastic energy per unit volume is
\begin{equation}
\phi(\epsilon_{ij})=\frac{1}{2} L_{ijkl} \epsilon_{ij} \epsilon_{kl}=\frac{1}{2}
\sigma_{ij} \epsilon_{ij}
\label{eqm3x}
\end{equation}

\begin{figure}[htb!]
\begin{center}
\resizebox*{100mm}{!}{\includegraphics{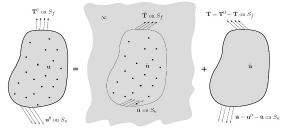}}
\end{center}
\caption{ Illustration of the superposition-based  boundary value
 problem solution for a body containing discrete defects. 
The $(\tilde{\ })$  fields correspond to the discrete defect fields in an
 infinite body and the $(\hat{\ })$ fields are the image fields that
 are obtained to satisfy the imposed boundary conditions. } 
\label{sketch}
\end{figure}

The superposition procedure for solving boundary value problems with a
collection of discrete defects introduced by \cite{Giessen95} is 
sketched in Fig.~\ref{sketch}. A collection of discrete defects is
assumed present in the solid. Each defect
is associated with am equilibrium  stress field and a displacement field. At any
time $t$, the stress $\tilde{\sigma}_{ij}(t)$ and displacement
$\tilde{u}_{i}(t) $ fields associated with the
collection of defects can be written as
\begin{subequations}
\begin{equation}
\tilde{\sigma}_{ij}(t) =\sum_{K=1}^N \sigma^{K}_{ij}(t)
\label{eqs1x}
\end{equation}
\begin{equation}
\tilde{u}_{i}(t) =\sum_{K=1}^N u^{K}_{i}(t)
\label{eqs1y}
\end{equation}
\end{subequations}
where $N$ is the number of discrete defects and for each defect
equilibrium stress field $\sigma^{K}_{ij,j}=0$. Although the stress
and deformation fields are associated with a solution to a linear
elastic boundary value problem, they are not necessarily
smooth. Indeed, in general, each displacement field $u^{K}_{i}(t)$ is
only piecewise differentiable and the corresponding stress field
$\sigma^{K}_{ij}(t)$ may be singular, as for a dislocation.

The $(\tilde{\ })$ stress and displacement fields are presumed to be
described by linear elasticity theory and  known either
from analytical expressions or from some separate semi-numerical
computation. The boundary conditions used to obtain the individual
defect fields are chosen 
for convenience, for example, for a defect in an infinite or
semi-infinite solid. Hence, the $\tilde{\sigma}_{ij}$ and
$\tilde{u}_i$ fields do not meet the imposed boundary conditions
\begin{equation}
T_i(t)=\sigma_{ij} n_j=T_i^0(t) \ { \rm on} \ S_T \quad , \quad
u_i(t)=u_i^0(t)  \ { \rm on} \ S_u
\label{bcs1}
\end{equation}
where $T_i^0(t)$ are the imposed traction vector components on $S_T$  and
$u_i^0(t)$ are the imposed displacement components on $S_u$. 

In order to meet the boundary conditions, the stress 
field $\hat{\sigma}_{ij}(t)$ and the displacement field
$\hat{u}_{i}(t)$ are introduced. The $(\hat{\ })$ fields are 
image fields and have no singularities 
in the body under consideration even if  $\tilde{\ }$ fields
do. The $(\hat{\ })$ fields are determined
from the solution to  the boundary value problem
\begin{equation}
\hat{\sigma}_{ij,j}=0
\label{eqbb1}
\end{equation}
with the boundary conditions
\begin{equation}
\hat{T}_i(t)=\hat{\sigma}_{ij} n_j=(T_i^0(t) -\tilde{T}_i )\ { \rm
  on} \ S_T \quad , \quad 
\hat{u}_i(t)=(u_i^0(t) - \tilde{u}_i)  \ { \rm on} \ S_u
\label{eqbb2}
\end{equation}
and $\hat{\sigma}_{ij}(t)=L_{ijkl} \hat{\epsilon}_{ij}(t)$.
Eqs.~(\ref{eqbb1}) and 
(\ref{eqbb2}) constitute a standard  linear elasticity problem that can
be solved numerically by a variety of methods, including the finite
element method, a boundary element 
method, a spectral method, etc. 

As illustrated in Fig.~\ref{sketch}, the total stress and displacement
fields are determined by superposition,
\begin{equation}
\sigma_{ij}(t)=\hat{\sigma}_{ij}(t) + \tilde{\sigma}_{ij}(t)
\label{eqs1}
\end{equation}
\begin{equation}
u_{i}(t)=\hat{u_{i}}(t)+ \tilde{u_{i}}(t)
\label{eqd1}
\end{equation}
and satisfy the governing equations and the imposed boundary
conditions. 

The $(\hat{\ })$ fields  generally vary more slowly spatially  than the
$(\tilde{\ })$ fields so that an advantage of this superposition
method is that in a numerical solution, a much coarser discretization
can be used than if the $(\tilde{\ })$ fields needed to be directly
solved for  in the calculation.

Note that in general it is not necessary for all defects to be of the
same kind but this superposition procedure can, in principle, be used
for some combination of dislocations, shear transformation zones,
vacancies, etc.

\subsection{Dissipation rate for discrete defect plasticity}

The Clausius-Duhem inequality for a purely mechanical formulation  is
expressed by Eq.~(\ref{cd12x}) and  the 
elastic energy per unit volume is given by Eq.~(\ref{eqm3x}).
Attention is confined to  circumstances
where the defect fields  are described by linear elasticity. The
boundary correction (or image) fields, $(\hat{\ })$, are continuous
and appropriately differentiable in $V$.  However,  the $(\tilde{\ })$
fields can have
surfaces of displacement/displacement rate discontinuity
as, for example, for 
dislocation fields and for \cite{Eshelby57,Eshelby59} transformation
fields. The displacement/displacement rate discontinuities are then
what give rise to  plastic deformation/plastic deformation rate.

Using a decomposition similar to 
Eq.~(\ref{eqs1}) for 
the strain rate $\dot{\epsilon}_{ij}$ 
$\dot{\tilde{\epsilon}}_{ij}$  is expressed in terms of the $K=1,2,...N$ individual 
defect strain rate fields as 
$$
 \dot{\tilde{\epsilon}}_{ij} = \sum_K \dot{\tilde{\epsilon}}^K_{ij} 
$$
 
and $\dot{W}$ is given by 
\begin{equation}
\dot{W}= \dot{W}_{\rm ext} +  \sum_{S^K}  \dot{W}^K= \int_{S_{\rm
    ext}} T_i \dot{u}_i dS +  \sum_{S^K}  \dot{W}^K  =\int_{V_e}
\sigma_{ij}  \left (\dot{\hat{\epsilon}}_{ij}  + 
\sum_K   \dot{\epsilon}^K_{ij}  \right ) dV   + \sum_{S^K}  \dot{W}^K
\label{eq4c2}
\end{equation}
where $T_i=\sigma_{ij} n_j$ are the traction vector components,
$S_{\rm ext}$ is 
the external surface with normal $n_j$, $\sigma_{ij}$ and
$\dot{u}_i$ are given by Eqs. (\ref{eqs1}) and (\ref{eqd1}),
respectively, $S^K$ is the surface of discontinuity  in $V$ associated
with  defect $K$, ${\dot  W}^K$ is the work rate on $S^K$,  
$\dot{W}_{\rm  ext}$ is the work rate on the external surface
of $V$ and $V_e$ 
denotes the volume of the material not including the 
$K=1,2,...N$ surfaces  $S^K$. 

The elastic energy rate per unit volume
is given by $\dot{\phi}=\sigma_{ij} \dot{\epsilon}_{ij}$ in the
material volume (which is where $\dot{\epsilon}_{ij}$ is defined) and
the dissipation rate for defect $K$ is the work 
rate $\dot{W}^K$ on its associated discontinuity surface $S^K$.
The elastic energy rate is given by
\begin{equation}
\dot{\Phi}=\int_{V_e} \dot{\phi} dV=\int_{V_e}\sigma_{ij} 
(\dot{\hat{\epsilon}}_{ij} +\dot{\tilde{\epsilon}}_{ij} )  dV
\label{eq4cc}
\end{equation}
Hence,
\begin{equation}
\dot{\cal D}=\dot{W}-\dot{\Phi}= \sum_K \dot{W}^K
\label{eq4cd}
\end{equation}

If compatibility (the strain/strain rate field can be derived from a continuous
differentiable displacement field) is satisfied pointwise there are no
surfaces of discontinuity and 
$\dot{\cal  D} =0$. Also, if a sub-volume of
$V$ is considered that does not contain a surface $S^K$, $\dot{\cal
  D}=0$ for that sub-volume. 

For defect $K$ with surface $S^K$ and with the tractions 
$T^K_i=\sigma_{ij} n^K_j$
($n^K_i$ is the normal to $S^K$) continuous across $S^K$, 
two possibilities for $\dot{W}^K$ are considered:
(i) in Eq.~(\ref{eq4c3}) the displacement rate jump $\Delta
\dot{u}^K_i$ evolves and $dS$ is fixed; 
and (ii)  in Eq.~(\ref{eq4c3z}) the surface $dS$ evolves and the
displacement jump $\Delta {u}^K_i$ is fixed. 

 If the surface $dS$ is fixed
\begin{equation}
\dot{\cal D}= \sum_K \dot{W}^K=\sum_{S^K} \int_{S^K} T^K_i \Delta \dot{u}^K_i dS 
\label{eq4c3}
\end{equation}
while if the displacement jump  $\Delta u^K_i $ is fixed 
\begin{equation}
\dot{\cal D} = \sum_K \dot{W}^K=\sum_{S^K} \int_{S^K} T^K_i \Delta {u}^K_i \dot{ dS}
\label{eq4c3z}
\end{equation}

\section{Discrete dislocation plasticity}

In deformation of structural metals, dislocation glide is typically
the main mechanism by which 
dislocations dissipate energy. A dislocation is characterized by its
Burgers vector ${\bf b}$,  its slip plane (the plane it is restricted
to glide on) and its glide direction. Dislocation glide dissipates
energy but  dislocations also store energy
due to their associated stress field. With the dislocations
represented as line defects in a linear elastic solid with a $1/r$
stress singularity, where $r$ is the distance from the dislocation
line, the elastic energy associated with a
dislocation varies as $\ln r$ and so approaches infinity both as $r
\rightarrow 0$ and $r \rightarrow \infty$. The latter is the more
interesting of these limits as it shows that the elastic energy of a
dislocation is not localized near the dislocation line. This has
important implications for dislocation interactions and pattern
formation, and, as a consequence, for the evolution of dissipation.

\subsection{Dissipation rate for discrete dislocation plasticity}

We consider a collection of $N$ dislocations with dislocation $K$
having a time independent displacement jump  specified by the Burgers
vector $b_i^K$ and gliding on a
slip plane with the components of the normal to the slip plane of
dislocation $K$ denoted by $n_i^K$.  The dissipation rate is
calculated from the expression for $\dot{W}^K$ 
associated with an infinitesimal  rigid motion of
dislocation $K$.  

As dislocation $K$  moves rigidly  along the glide plane, the
rate of change of $dS$ is  $v^K dl$  where $dl$ is an element of the
dislocation line  and $v^K$ is 
the dislocation velocity normal to $dl$, so that for dislocation $K$
Eq.~(\ref{eq4c3z}) gives 
\begin{equation}
T_i \Delta {u}^K_i \dot{dS}= \left (
\hat{\sigma}_{ij}+ \sum_{M=1,\ne K}^N
\sigma^{M}_{ij}  \right )  n^{K}_j  b^{K}_i v^{K} dl
\label{eqd0a}
\end{equation}
  where $\hat{\sigma}_{ij}$ is the image stress field in
  Eq.~(\ref{eqs1}) and
  $\sigma^{K}_{ij}$ does not contribute to $\dot{W}^K$  because
it does not induce a change in energy for an  infinitesimal rigid motion
of dislocation $K$. 
 
The total dissipation rate can then be written as
\begin{equation}
\dot{\cal{D}}=\sum_{K=1}^N \int_{L^{K}} P^{K} v^{K} dl
\label{eqd1a}
\end{equation}
where $P^K$, the Peach-Koehler (configurational) force is given by
\begin{equation}
P^{K}=\left ( \hat{\sigma}_{ij} +  \sum_{M=1,\ne K}^N
\sigma^{M}_{ij} \right )  n^{K}_j  b^{K}_i
=  \left ( \hat{\sigma}_{nb} +  \sum_{M=1,\ne K}^N
\sigma^{M}_{nb} \right ) \vert \mathbf{b}^K \vert
\label{eqd2a}
\end{equation}
where   $\sigma_{nb} $ is the Schmid resolved shear stress given by
$\sigma_{nb} = \bar{b}^{K}_i \sigma_{ij}  \, n^{K}_j$ with 
$\bar{b}^{K}_i$ a unit vector in the Burgers vector direction of
dislocation $K$ and
$\vert \mathbf{b}^K \vert$ the Burgers vector amplitude. 
A detailed direct calculation of the change in elastic energy due to
 dislocation glide leading to the expression Eq.~(\ref{eqd2a}) is
 given by \cite{Lubarda93}.  

A kinetic relation for $v^K$ needs to be specified. For modeling 
crystals such as fcc metal crystals, a linear relation of the
form 
\begin{equation}
v^{K}=\frac{1}{B^{K}} P^{K} 
\label{eqd3aa}
\end{equation}
is typically used   where $B^K$ is termed the dislocation
  mobility. Eq.~(\ref{eqd3aa}) implies that plastic flow in 
such a crystal only depends on the shear stress in the slip plane
in the Burgers vector direction,   i.e. the Schmid resolved shear stress.
However, there are circumstances 
where there is evidence from atomistic calculations and experiment
that other shear stress components can play a role;  for
example,  in modeling dislocation glide when cross-slip occurs,
e.g. \cite{Huss15,Malka21}, and   in  modeling bcc crystals as well as
crystals that have a  complex dislocation core structure so that 
shear stress components other than $\sigma_{nb}$ can affect
dislocation mobility,  e.g. \cite{Bass92,Vitek04,Vitek08,Wang11}. In
such cases a possible kinetic relation  that has non-Schmid shear stress
dependence is of the form 
\begin{equation}
v^{K}=\frac{1}{B^{K}} Q^{K} = \frac{1}{B^{K}} \left [ P^K + \rho^K \left
  (  \hat{\sigma}_{mt} + 
  \sum_{M=1,\ne K}^N c_M \sigma^{M}_{mt} \right ) \right ]
\label{eqd3a}
\end{equation}
where $\sigma_{mt} = m^{K}_i \sigma_{ij}  \, t^{K}_j$with $m^{K}_i$
and $t^{K}_j$ unit vectors in directions  that differ from $n_i$
and/or  $\bar{b}_i$.  

In other circumstances, a  possibility is that the dislocation
velocity depends on the 
hydrostatic stress so that, for example, $Q^K$ is given by an
expression such as 
\begin{equation}
v^{K}=\frac{1}{B^{K}} Q^{K} = \frac{1}{B^{K}}  \left [ P^K + \rho^K
  \sigma_{kk} \right ]
\label{eqd3b}
\end{equation}

In either case, the dissipation rate is 
\begin{equation}
\dot{\cal{D}}=\sum_{K=1}^N \int_{L^{K}} \left ( \frac{1}{B^{K}}
\right ) P^{K} Q^{K}  dl =\sum_{K=1}^N \dot{\cal{D}}^K
\label{eqd1c}
\end{equation}
with $B^{K}\ge 0$ and $Q^{K} =P^{K}$, $\dot{\cal{D}} $ in
Eq.~(\ref{eqd1a}) is guaranteed to be non-negative. However, with 
$Q^{K} \ne P^{K}$ there is in general no guarantee that $P^{K} Q^{K}
\ge 0$ for  all $K$ and all possible stress states.  

The expression for $\dot{\cal{D}}$
in Eq.~(\ref{eqd1c}) pertains to changes in the
dislocation positions. Any contribution of a change  in
the lengths of dislocation lines to a change in
stored elastic energy  and/or to a change in dissipation is not
accounted for by Eq.~(\ref{eqd1c}). 

\subsection{Calculations of discrete dislocation plasticity
  dissipation}
\label{calc-disl}

The evolution of the dissipation rate, Eq.~(\ref{eqd1c}),  depends
on many factors in addition to the dislocation parameters, including
the material microstructure and the imposed loading history, as well
as any relevant geometric  size scales. Examples that illustrate such
dependencies are presented  for single crystals using two-dimensional
plane strain analyses that only involve  
edge dislocations. In all cases the dislocation velocity is
proportional to the Peach-Koehler force,  $P^{K}= Q^{K}$ and
$B^K>0$ in 
Eq.~(\ref{eqd1c}), so that a non-negative glide dissipation rate is
guaranteed. The presentation here is limited to qualitative features
of the response, details of the formulations and parameter values used 
in the calculations are given in the references cited. 
Also, although the general formulation is 3D, the examples discussed
are 2D plane strain.

\begin{figure}[htb!]
\begin{center}
{\resizebox*{50mm}{!}{\includegraphics{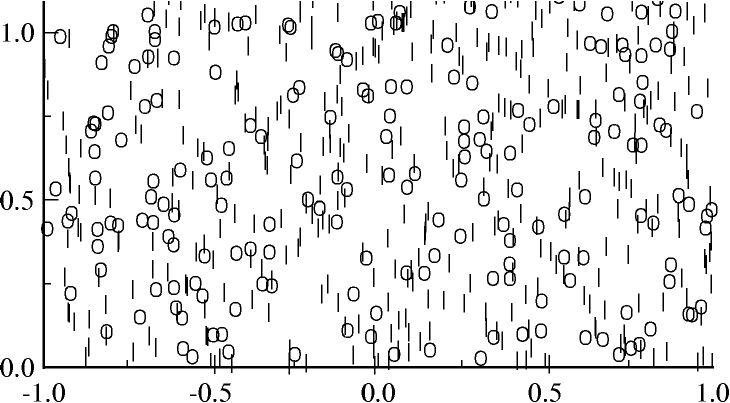}}}
\end{center}
\caption{A random distribution of dislocation sources, circles, and
  dislocation obstacles, lines.}
\label{chi1b}
\end{figure}

The two-dimensional crystals considered are initially 
free of mobile dislocations  and a  random  
distribution of dislocation sources and a random  
distribution of  dislocation obstacles are specified as illustrated
in Fig.~\ref{chi1b}. 
A Gaussian distribution dislocation  source strengths is
specified with a mean Peach-Koehler force for nucleation and an
associated standard deviation. Similarly, a Gaussian distribution of
dislocation obstacle strengths is specified.  The same  dislocation
mobility constant $B^K$ is prescribed for all dislocations. 

\begin{figure}[htb!]
\begin{center}
\subfigure[]
{\resizebox*{65mm}{!}{\includegraphics{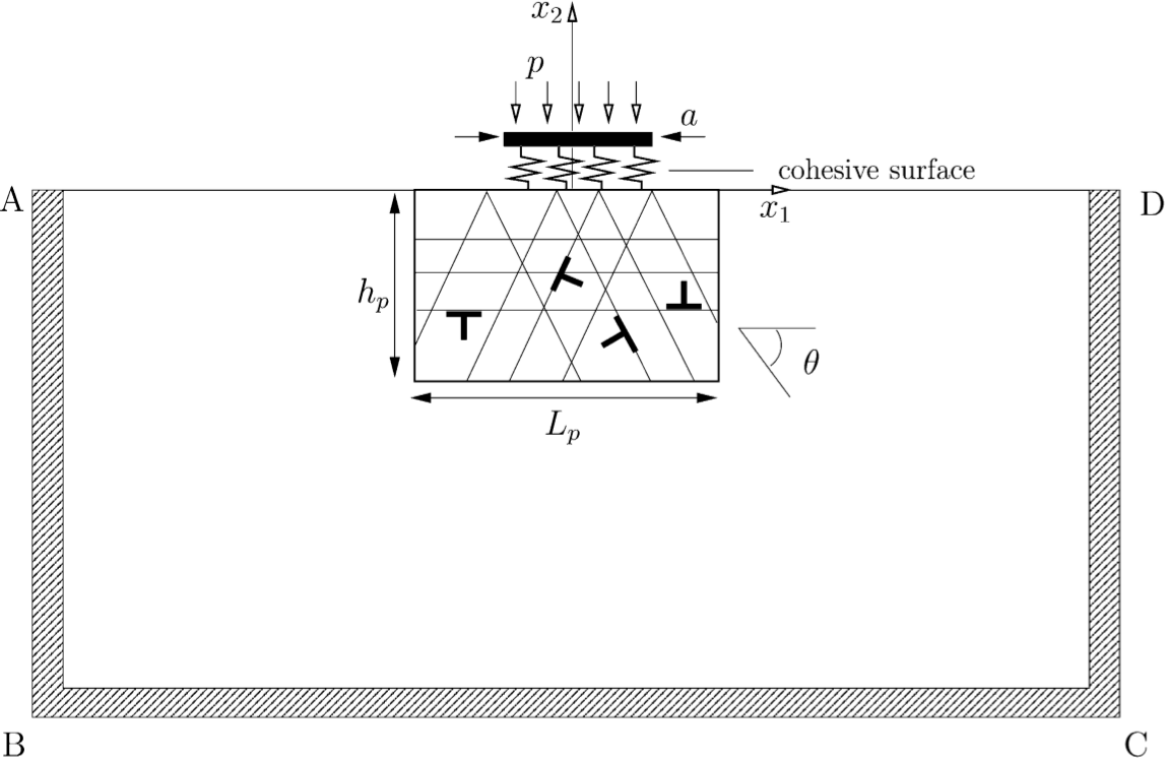}}}
\subfigure[]
{\resizebox*{65mm}{!}{\includegraphics{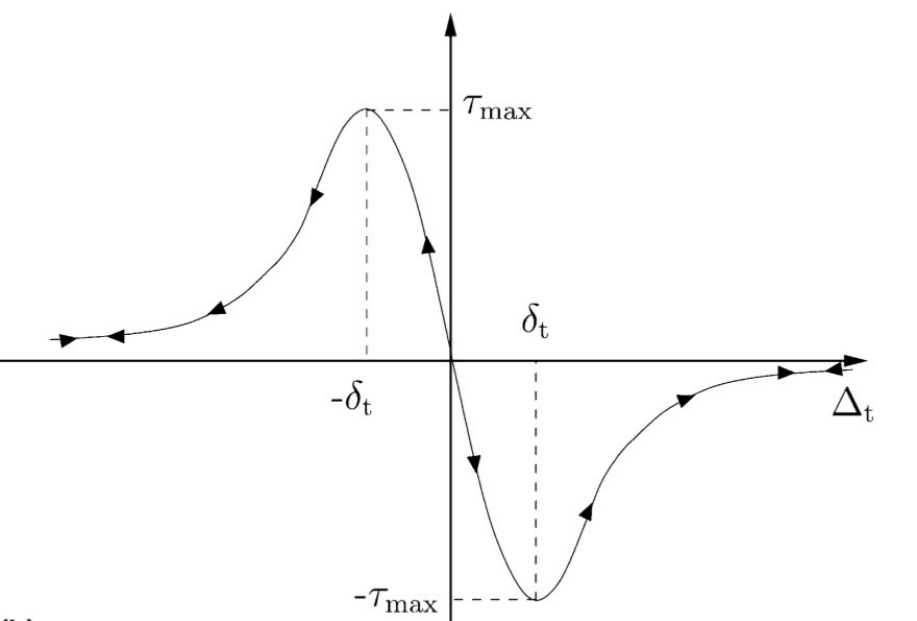}}}
\end{center}
\caption{ (a) Sketch of the boundary value problem to model the initiation
  of sliding by \cite{fric04,fric05}. (b) The softening shear cohesive
  relation used by \cite{fric04,fric05} .}
\label{slid}
\end{figure}

\begin{figure}[htb!]
\begin{center}
\subfigure[]
{\resizebox*{70mm}{!}{\includegraphics{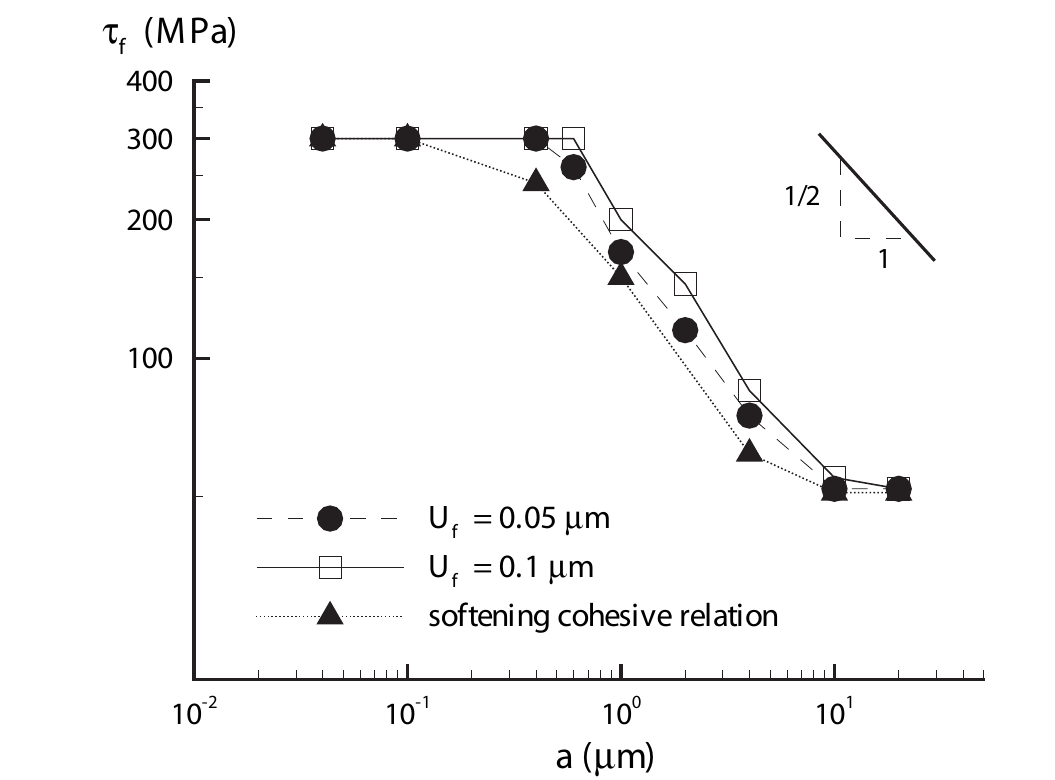}}}
\subfigure[]
{\resizebox*{70mm}{!}{\includegraphics{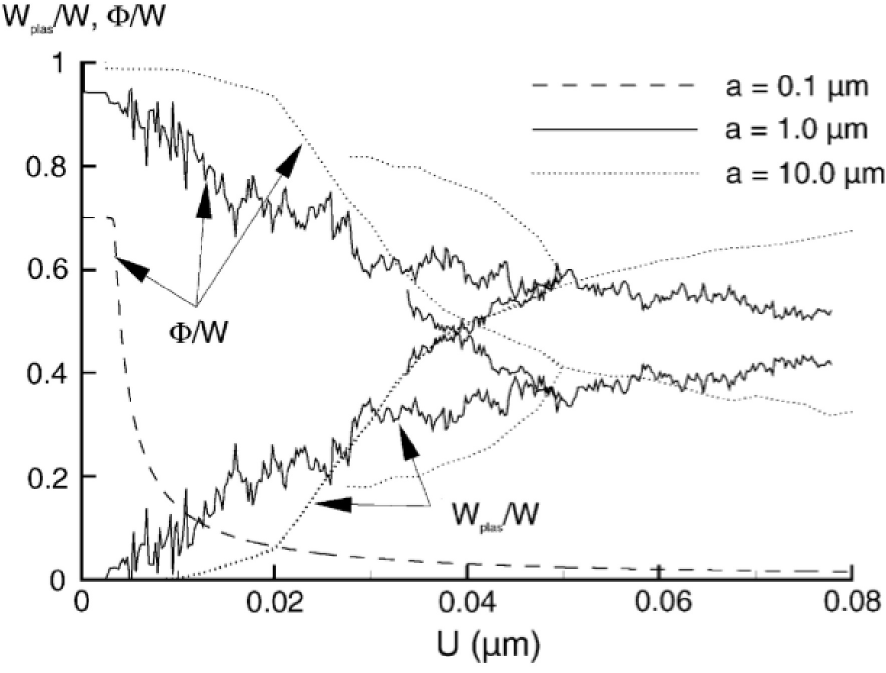}}}
\end{center}
\caption{(a)Friction stress as a function of contact size $a$. 
(b)   Plastic dissipation $W_{\rm plas}$, strain energy
  $\Phi$ and cohesive energy $W_{\rm cohes}$ normalized by the work
  $W$ as a function of sliding displacement $U$. From \cite{fric04,fric05}}
\label{frict-s}
\end{figure}

\cite{fric04,fric05} carried out calculations modeling the initiation
of sliding over asperities of various sizes. 
The plane strain boundary value problem sketched in 
Fig.~\ref{slid}a  was analyzed. Monotonically
increasing displacements were applied on the boundaries AB, BC and CD
in  the $x_1-$direction to give an overall shear displacement $U$
together  with a normal pressure of magnitude $p$ over 
the asperity length $a$.  Two types of shear cohesive relations (a
cohesive relation specifies the relation between the traction/traction
rate across
an interface and the displacement/displacement rate jump across it) were
used to characterize the interface: (i) a softening cohesive relation,
i.e. the shear stress decreases to zero for a sufficiently large
relative displacement as shown in Fig.~\ref{slid}b; and (ii) a
non-softening relation in which the shear 
traction reaches a maximum and then remains constant.
For the softening cohesive relation,
the  initiation of sliding was identified with the shear traction
reaching zero within a numerical tolerance. For the non-softening
cohesive relation the initiation 
of sliding was identified with attaining a specified value of
shear displacement.   Also, as sketched in Fig.~\ref{slid}, dislocation
activity was restricted to a region around the contact. 
Calculations were terminated prior to dislocations reaching the
boundary of that region. 

The definition of the friction stress $\tau_{\rm fr}$ depended on the
shear cohesive relation used. For the softening cohesive relation, it
was the shear stress just before the shear stress magnitude dropped to
zero. For 
the non-softening cohesive relation, $\tau_{\rm fr}$  was the shear
stress at the value  of shear displacement $U$ identified  as the
initiation of 
sliding. Thus, for the non-softening cohesive relation the value of
$\tau_{\rm fr}$ was somewhat arbitrary. 

Fig.~\ref{frict-s}a shows the effect of the contact size $a$ on
$\tau_{\rm fr}$. For the non-softening cohesive relation, results are
shown using two values of shear displacement $U$ to define the onset
of sliding that 
differ by a factor of two. Regardless of the definition of sliding
initiation, 
the variation of the shear stress at the initiation of sliding,
$\tau_{\rm fr}$, with contact size $a$ 
exhibits two plateaus: for large contacts $\tau_{\rm fr}$ was found to
be  approximately equal to the
tensile yield strength, while for small contacts, it is 
equal to the cohesive strength,  which is $\approx 6$ times larger
than the yield strength. In between these two plateaus, $\tau_{\rm
  fr}$ varies as $\sqrt{a}$.

The evolution of  plastic dissipation,   i.e. $\int {\cal D} dt$
  and denoted by $W_{\rm plas}$
  in \cite{fric04,fric05},  elastic  energy, 
$\Phi$, and cohesive energy, $W_{\rm cohes}$ (the energy stored in the
cohesive surface), are shown in 
Fig.~\ref{frict-s}b for $a=0.1\mu$m (the upper plateau), $1.0\mu$m
(the transition region) and $10.0\mu$m (the lower plateau). On the upper
plateau, $a=0.1\mu$m , the energy is partitioned into elastic energy $\Phi$
and cohesive 
energy $W_{\rm cohes}$, with the elastic energy dominating the early
stages of sliding 
and the cohesive energy the latter stages of sliding. On the lower
plateau, $a=10.0\mu$m, the partitioning is mainly between $\Phi$ and
$W_{\rm plas}$. In between, $a=1.0\mu$m, all three components of the
energy are significant. This illustrates the predicted  contact size
dependence of  energy partitioning for the initiation of frictional
sliding. 

\begin{figure}[htb!]
\begin{center}
\resizebox*{90mm}{!}{\includegraphics{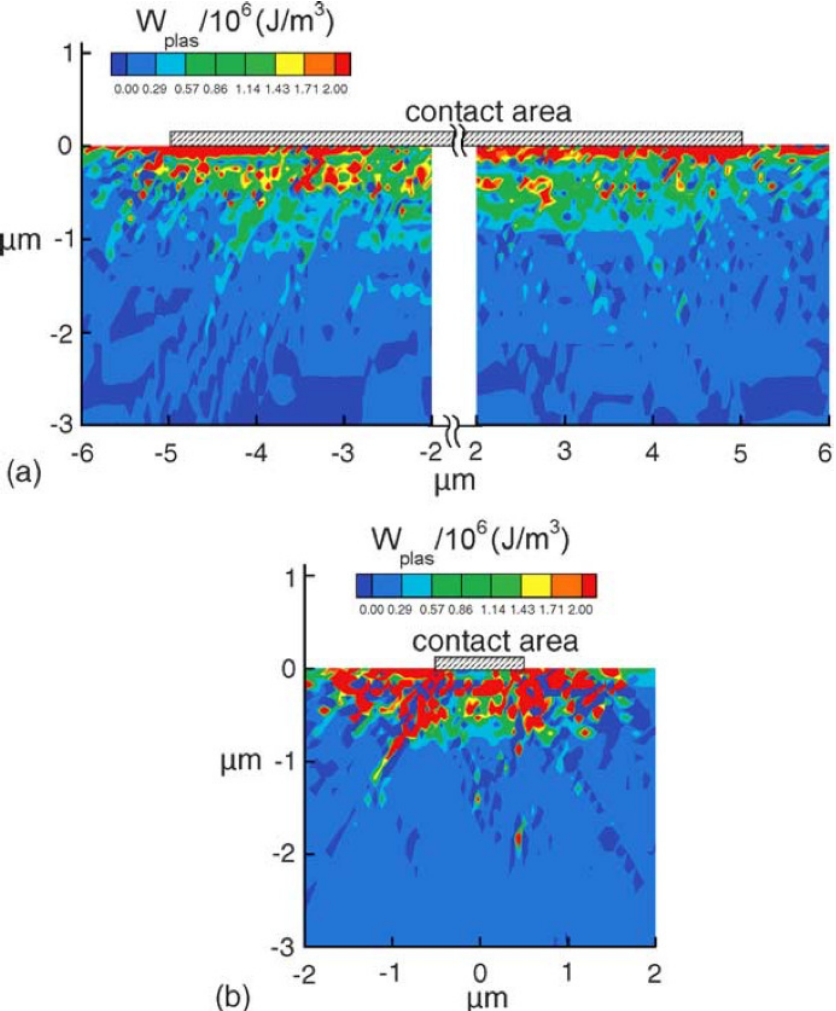}}
\end{center}
\caption{Distribution of the plastic dissipation $w_{\rm plas}$ per unit
volume for: (a) contact size $a=10\mu$m and (b) contact size 
$a=1\mu$m. The contact area is indicated. From \cite{fric05}.}
\label{frict-e}
\end{figure}

Fig.~\ref{frict-e}  shows the distribution of  plastic dissipation per
unit volume, $W_{\rm plas}$, for the $a=1.0\mu$m  and $a=10.0\mu$m
sizes.  For $a=10\mu$m,  plastic dissipation mainly occurs near the
contact surface but the  high dissipation region extends somewhat in front
of and behind the contact area. For $a=1.0\mu$, the high dissipation
region extends much more deeply into the crystal. Thus, not only the
energy partitioning but also the spatial distribution of plastic
dissipation is size dependent.

\begin{figure}[htb!]
\begin{center}
\subfigure[]
{\resizebox*{25mm}{!}{\includegraphics{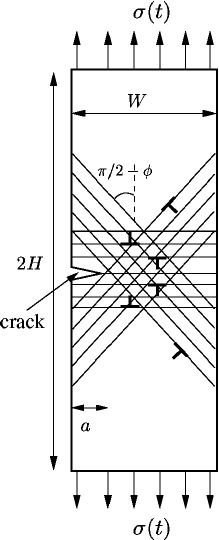}}} \qquad
\subfigure[]
{\resizebox*{50mm}{!}{\includegraphics{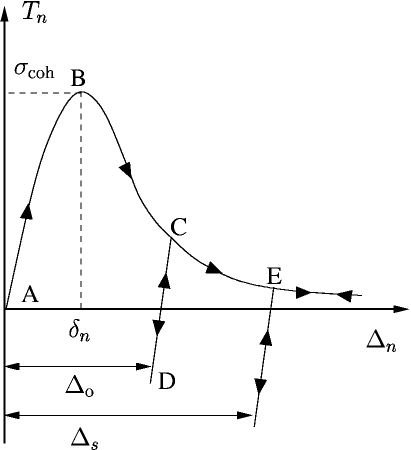}}}
\end{center}
\caption{(a) Sketch of the single crystal short crack boundary value problem
  analyzed by \cite{short03}. (b) The cohesive
  relation used by \cite{short03}.}
\label{short1}
\end{figure}

The evolution of dissipation under cyclic loading, i.e. hysteresis, 
 plays a major role in determining the fatigue behavior
of materials.  It has long been appreciated that short fatigue cracks
in metallic components grow faster 
than long cracks subject to the same cyclic loading
range. For sufficiently long cracks, fatigue crack growth is governed by a critical
stress intensity factor  range (the stress intensity factor $K$ is
proportional to $\sigma \sqrt{a}$, where $\sigma$ is the applied stress
and $a$ is the crack length), while for
sufficiently short cracks, fatigue crack growth is governed by a
critical stress range, see e.g. \cite{Suresh}.

\begin{figure}[htb!]
\begin{center}
\subfigure[]
{\resizebox*{70mm}{!}{\includegraphics{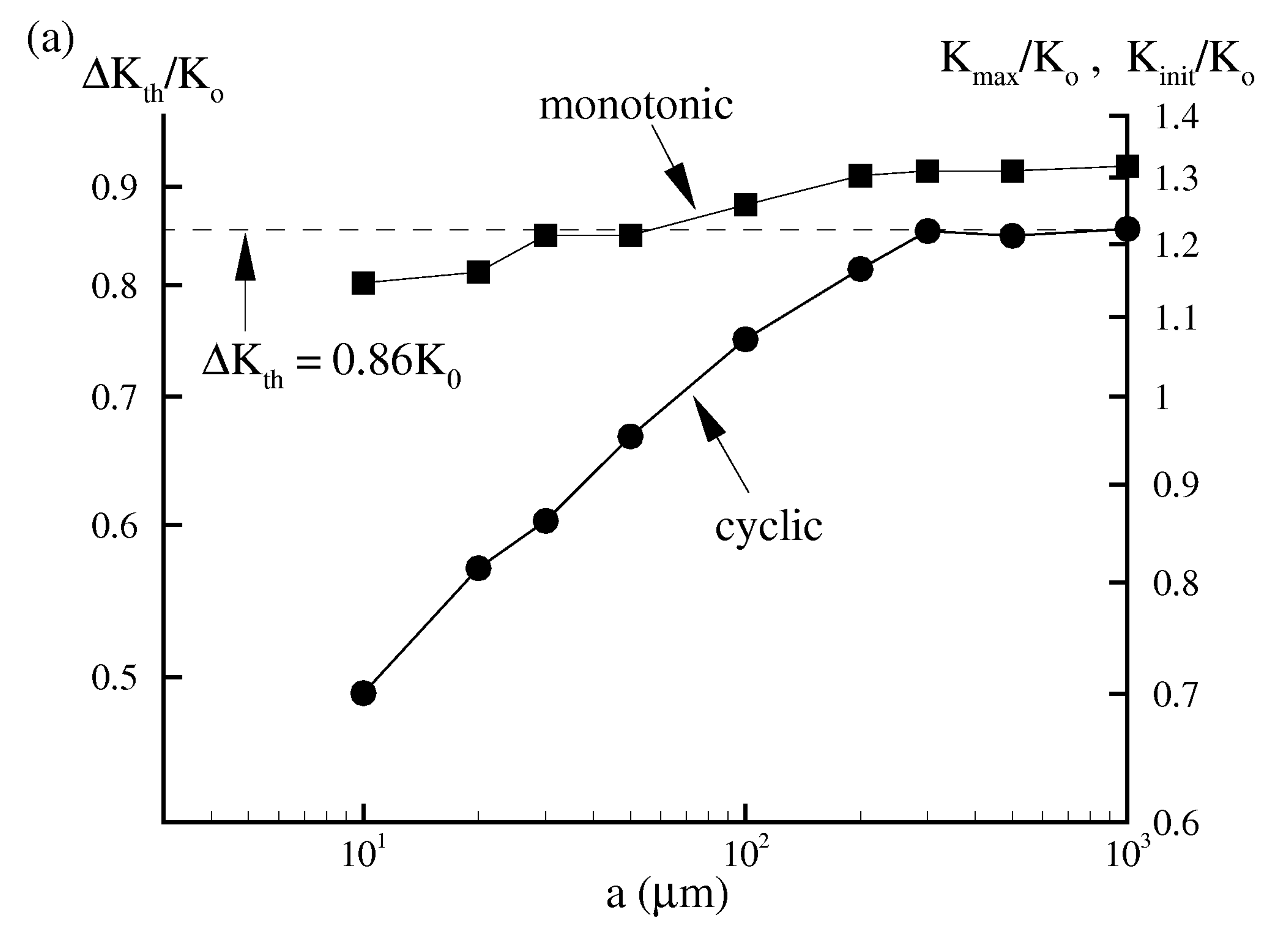}}}
\subfigure[]
{\resizebox*{60mm}{!}{\includegraphics{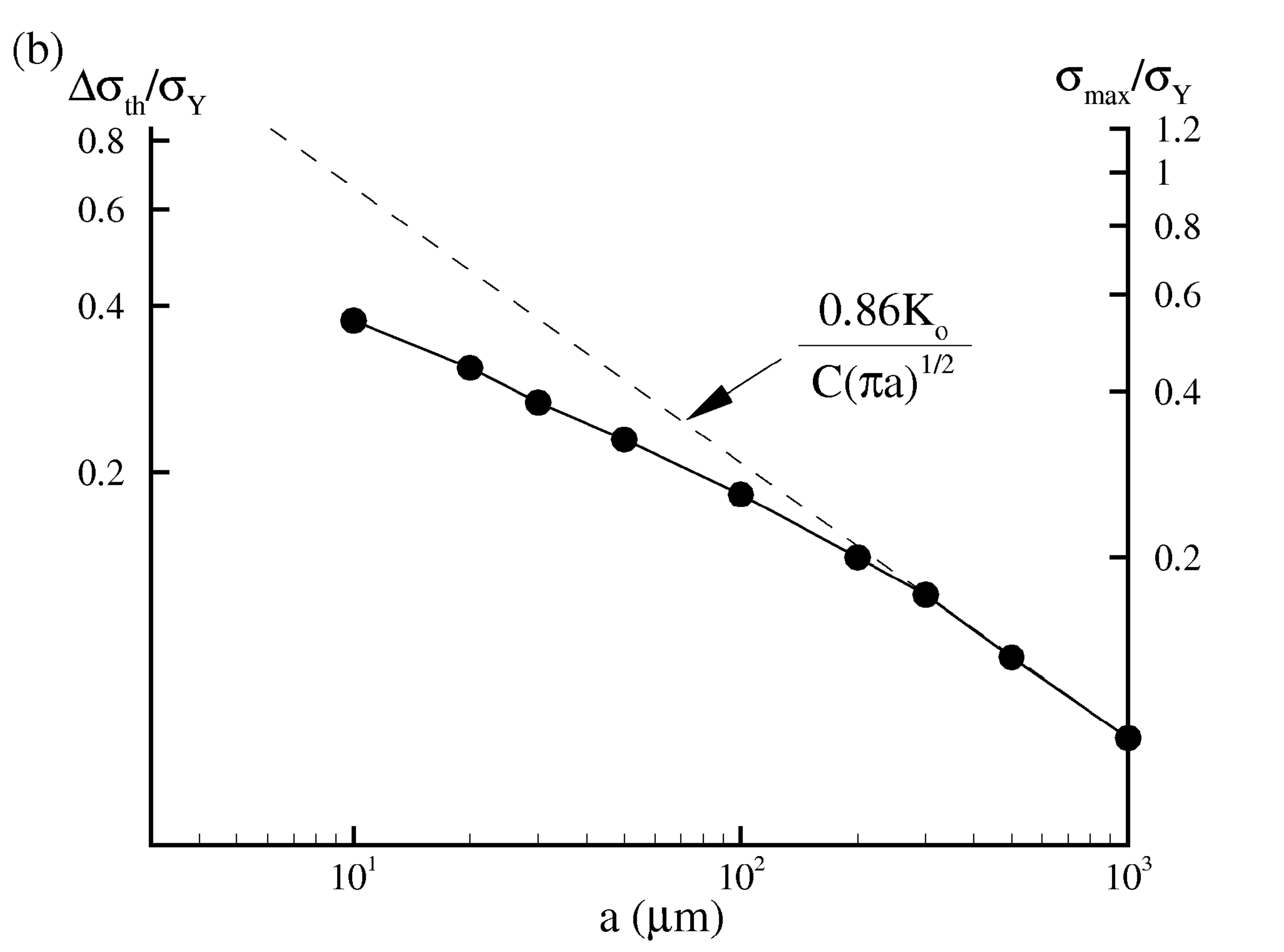}}}
\end{center}
\caption{(a) The normalized fatigue threshold  $\Delta K_{\rm th}/K_o$
   versus initial crack length $a$.  (b) The normalized fatigue
 threshold  versus crack length $a$  as a function of  $\Delta
 \sigma/\sigma_y$ (left axis) and as a function of $\sigma_{\rm
  max}/\sigma_y$  (right axis). The dotted
lines show the predictions for  $\Delta K$ governed crack growth.
From \cite{short03}.}
\label{short2}
\end{figure}

\cite{short03} carried out calculations of
crack growth for 
various edge cracked single crystals having the geometry sketched in 
Fig.~\ref{short1}a (in the calculations of \cite{short03}, as in
those of \cite{fric04,fric05} dislocation activity was confined to a
region near the crack tip). The presumption underlying the analyses of 
\cite{short03} is that crack growth is stress driven and that
the main source of dissipation is a consequence of the glide of
dislocations that nucleate from sources in the crystal. Stress driven
crack growth is modeled via the tensile cohesive relation shown
in Fig.~\ref{short1}b. 
Two applied tensile displacement 
 loading histories were considered: (i) displacements that
 monotonically increase; and  (ii) displacements that are a cyclic
 function of time.  

The results of  \cite{short03} 
are summarized in Fig.~\ref{short2}a which shows the dependence of the
predicted initiation of crack growth 
under both   monotonic and cyclic loading for geometrically
similar specimens as a function of  edge crack length $a$. 
For the edge cracked specimen the relation between the overall stress
$\sigma$ and the linear elastic stress intensity factor $K$ is
known. For a sufficiently long crack, the linear 
elastic stress intensity factor acts as the 
crack driving force  and cyclic loading is 
characterized by $\Delta K=K_{\rm max}-K_{\rm min}$. The fatigue
threshold $\Delta K_{\rm th}$ is the smallest value of $\Delta K$ that
is found to lead to crack growth (in general $\Delta K_{\rm th}$
depends on the ratio $K_{\rm min}/K_{\rm max}$). Also, the
Griffith stress intensity factor, $K_0$, is proportional to the square
root of the cohesive energy. 

For monotonic loading
$K_{\rm init}/K_0$ is shown (right axis) and for cyclic loading both
$\Delta K_{\rm th}/K_0$  (left axis) and $K_{\rm max} /K_0$ (right
  axis) are shown. In Fig.~\ref{short2}b the fatigue threshold results for
  cyclic loading are plotted versus $\Delta \sigma=\sigma_{\rm max}
  -\sigma_{\rm min}$ and $\sigma_{\rm max}$ showing the transition to
  stress controlled, rather than stress intensity controlled, crack
  growth resistance for sufficiently short cracks. 

Discrete dislocation plasticity analyses predict that organized
dislocation structures that develop in the vicinity of a crack  tip
can result in crack tip opening stresses  approximately an
order of magnitude larger than predicted by conventional continuum
plasticity, e.g. \cite{Clever}. These near crack tip  organized
dislocation structures increase stress levels and hence the associated
defect stored elastic energy.

The results of \cite{short03} indicate
that the development of the near crack tip  stress fields and their
associated dissipation  depends on both the crack length and whether
the loading is monotonic or cyclic. 
Although both the monotonic and cyclic 
crack growth results exhibit a decreased resistance to crack growth
for sufficiently short 
cracks, the decline in crack growth resistance for short cracks is
much greater under 
cyclic loading conditions than it is under monotonic loading conditions.
For short cracks under cyclic loading conditions, 
crack growth was found to take place for values of 
$K_{\rm max} < K_0$. 

\begin{figure}[htb!]
\begin{center}
{\resizebox*{80mm}{!}{\includegraphics{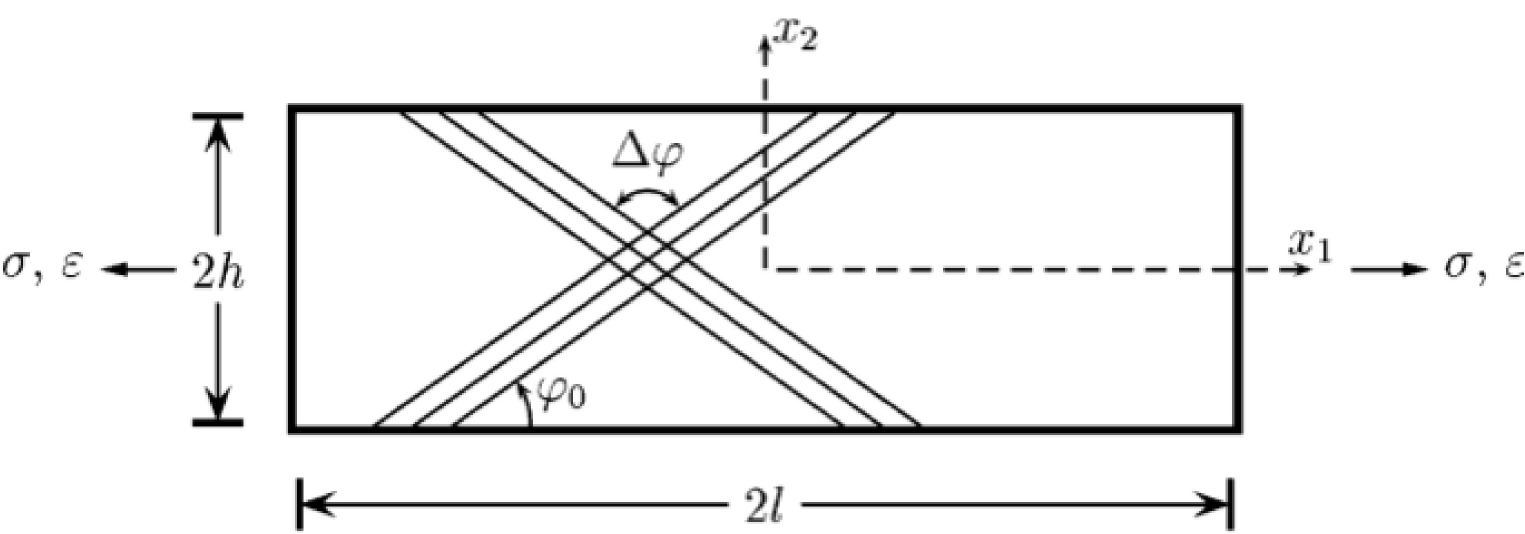}}}
\end{center}
\caption{The plane strain single crystal analyzed by \cite{Amine05}. }
\label{chi1}
\end{figure}

\begin{figure}[htb!]
\begin{center}
\subfigure[]
{\resizebox*{70mm}{!}{\includegraphics{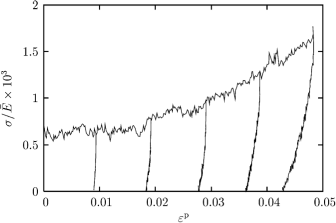}}}
\subfigure[]
{\resizebox*{70mm}{!}{\includegraphics{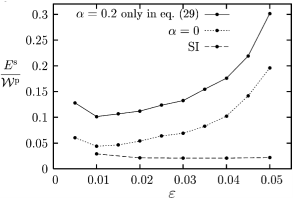}}}
\end{center}
\caption{(a) The normalized stress strain curve with unloading showing
the predicted Bauschinger effect. (b)  Ratio of the stored 
energy, $E_s$, to plastic 
work, $W_p$, versus imposed strain, $\epsilon$.   The stored energy
$E_s$ is defined as the elastic energy $\phi$ minus the elastic energy
associated with the overall stress. Also,  $\alpha$ is a
parameter characterizing the dislocation line tension and SI denotes
the response with only static sources and obstacles. From \cite{Amine05}.} 
\label{chi2}
\end{figure}

\begin{figure}[htb!]
\begin{center}
\resizebox*{90mm}{!}{\includegraphics{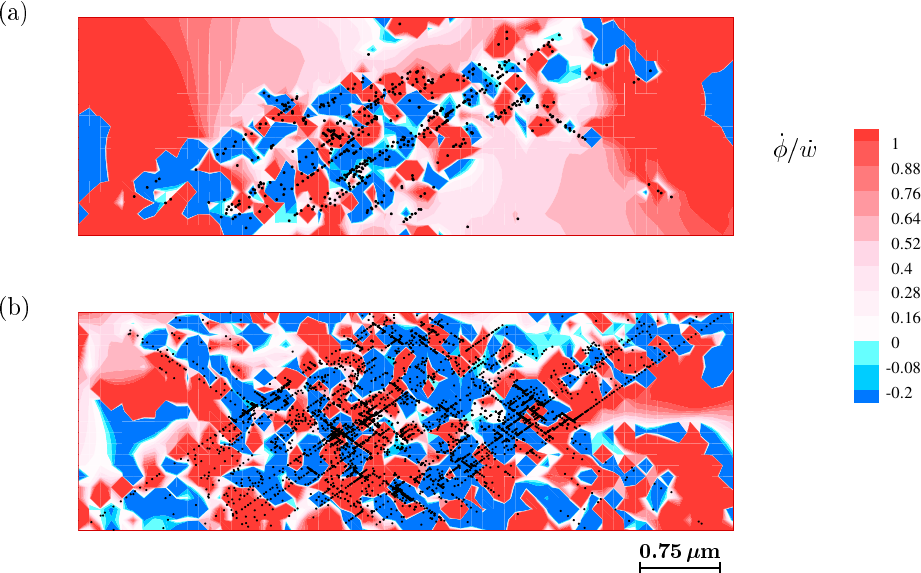}}
\end{center}
\caption{Distributions of the ratio of the local stored energy
  rate,$\dot{\phi}$, to the local rate of working, $\dot{w}$. (a)
  $\epsilon=0.01$. (b) $\epsilon=0.04$. From \cite{Amine05}.} 
\label{chi4}
\end{figure}

Knowledge of the partitioning between stored elastic energy and
dissipation is needed for predicting the  deformation modes
that  occur in machining and penetration, and for predicting 
 the thermal softening
behavior that promotes mechanical instabilities, Furthermore,  the
stored energy is associated the internal stress state so that there is
a direct connection between the energy partitioning 
and the Bauschinger effect. 

\cite{Amine05} calculated the  energy partitioning for a single
crystal under plane strain tensile loading. Fig.~\ref{chi1}
illustrates the plane strain tension specimen 
analyzed.  A major difference between
the plane strain tension problem and those modeling sliding and crack
growth is that there are no imposed strain gradients for plane strain
tension.  With no imposed strain gradient,  the simple constitutive
rules used in those 
analyses do not give rise to dislocation structures that contribute to
the elastic energy. In \cite{Amine05} the two-dimensional discrete
dislocation 
plasticity constitutive rules used to model  sliding and crack
growth were augmented to account for the
effects of line tension,  three-dimensional dislocation
interactions and, in particular, the dynamic creation of new obstacles
and sources  as proposed by \cite{Amine03}. 

One crystal geometry analyzed had slip systems with an initial
unsymmetrical orientation relative to the loading axis. 
Fig.~\ref{chi2} shows results for such a crystal. In Fig.~\ref{chi2}a
 there is a strong Bauschinger effect, the magnitude of which increases
 with increasing strain. 
Fig.~\ref{chi2}b shows the
evolution of the ratio of stored energy $E^s$ to plastic work
$W^p$. The parameter $\alpha$ 
characterizes line tension with 
$\alpha=0$ denoting that line tension is not included in the
analysis. The  result marked SI denotes a calculation with only static
sources and obstacles that results in 
the energy storage being very small $E^s/W^p < \approx 0.04$, so that
nearly all plastic working is dissipated. When the dynamic creation of
sources and obstacles is included in the analysis, 
significantly more elastic energy is stored.
The quantity $1-E^s/W^p$ is the fraction of plastic
work dissipated that is used in phenomenological calculations to induce
thermal softening. The modeling results in Fig.~\ref{chi2}b and
the experimental results of \cite{Hod00} indicate that this fraction is
not necessarily a constant but can vary with the loading.

Fig.~\ref{chi4} shows the distribution of the rate of energy
storage,$\dot{\phi}/\dot{w}$, 
and the dislocation positions at two values of overall strain
$\epsilon$.
In regions that remain essentially elastic,  $\dot{\phi}/\dot{w} \approx 1$, 
as expected.  In regions with a high dislocation
density this ratio has peak values much larger than the overall
ratio, $\dot{\Phi}/\dot{W}$ showing that   local values of the energy
storage rate can differ significantly from the overall global value.  

A beginning of the emergence of organized dislocation structures can
be seen in Fig.~\ref{chi4}.  Such organized dislocation structures
affect the internal stress field and, hence, the energy
partitioning. Without a strong applied stress/strain gradient to drive
the dislocation structure formation, the details of dislocation
interactions dominate. Most likely, a full three-dimensional discrete
dislocation plasticity analysis is needed to provide an accurate
picture of their development. However, such dislocation structures
develop at moderate to large deformations and nearly all discrete
dislocation plasticity analyses have been based on small deformation
kinematics which, among other limitations, cannot reasonably account
for the effects of
lattice rotations. Finite deformation formulations are available,
\cite{finit1,finit2},  but use has been very limited. 

For both the Bauschinger effect in Fig.~\ref{chi2}a and
the short crack effect in Fig.~\ref{short2},  a key role is played by
the evolution of the discrete dislocation structure 
under reverse loading in the 
heterogeneous internal stress state induced by the discrete
dislocation structure that develops under  forward loading. Indeed,
the  release of stored elastic energy  associated with such a stress state 
plays a key role in driving fatigue crack growth with 
$K_{\rm max} < K_0$.

\section{Discrete STZ plasticity}

Plastic deformation in a variety of amorphous solids occurs via
discrete regions of atomic rearrangement called  shear transformation
zones (STZs), \cite{Argon79, sp77} and the representation of STZs in
terms of transforming \cite{Eshelby57,Eshelby59} inclusions was pioneered by
\cite{Bulatov94}. 

A transforming  \cite{Eshelby57,Eshelby59} inclusion is an 
 ellipsoidal  sub-volume   of a uniform linear 
elastic solid undergoes a stress-free transformation, i.e. if the
transforming inclusion had been removed 
from the matrix, it would have undergone a uniform strain with zero
stress. Since the inclusion was not free to deform, it and the
surrounding matrix are stressed.  For an  ellipsoidal inclusion, the
strain and stress fields inside the inclusion associated with the
constrained transformation are also uniform. Due to the displacement
field associated with the transformation  there is  a
displacement jump  across the
inclusion/matrix interface. In general, the transformation reduces the stress
level inside the inclusion and induces a stress concentration outside
the inclusion.  

Transforming  \cite{Eshelby57,Eshelby59} inclusions are used to model
a wide variety of phenomena in addition to STZs including, for
example, phase 
transformations and internal stress states in heterogeneous
materials.  For discrete STZ plasticity, the location and size of an 
STZ is fixed, and the transformation strain evolves. Also, for
discrete STZ plasticity, the stress field outside the inclusion varies
as $1/r^2$ (2D) or $1/r^3$ (3D) which gives a much more rapid stress
decay than the $1/r$ variation for dislocations. Other applications,
for example the modeling of some phase transformations, may 
involve  \cite{Eshelby57,Eshelby59} 
inclusions with a size that varies with time and with either a fixed or
evolving transformation strain. 
Although the focus here is on shear transformations,
the general results regarding dissipation rate apply
to  fixed-size
\cite{Eshelby57,Eshelby59}  inclusions with general transformation
strains. 

\subsection{Dissipation rate for discrete STZ plasticity}

In discrete STZ plasticity modeling, $dS$ in Eq.~(\ref{eq4c3}) is fixed and
the displacement rate jump evolves. With the transformation strain
rate of inclusion $K$ 
denoted by 
$\dot{\epsilon}^{*K}_{ij}$, the displacement rate jump across 
interface $K$ is 
 $ \Delta \dot{u}^{K}_i = \dot{\epsilon}^{*K}_{im} x_m$ in Eq.~(\ref{eq4c3}) so that
\begin{equation}
\dot{\cal D}= \sum_{K=1}^N\int_{S^{K}}\sigma^K_{ij} n_j \dot{\epsilon}^{*K}_{im} x_mdS
\end{equation}
or changing to a volume integral for an \cite{Eshelby57,Eshelby59}
inclusion
\begin{equation}
\dot{\cal D}= \sum_{K=1}^N\left [\int_{V^{K}}\sigma^K_{ij}
dV  \right ]  \dot{\epsilon}^{*K}_{ij} +  \sum_{K=1}^N\left
    [\int_{V^{K}}\sigma^K_{ij,j}   x_m dV \right ]
    \dot{\epsilon}^{*K}_{im} 
\label{ent00}
\end{equation}
and because $\sigma^K_{ij,j} =0$ 
\begin{equation}
\dot{\cal D} =  \sum_{K=1}^N\left [ \int_{V^{K}}
  \left ( \hat{\sigma}_{ij} +\sum_{M=1}^N 
\sigma^{T_M}_{ij} \right ) dV \right ] \dot{\epsilon}^{*K}_{ij}
=\sum_{K=1}^N Z^K_{ij} \dot{\epsilon}^{*K}_{ij} 
\label{eq7zz}
\end{equation}
where $\sigma^{T_M}_{ij} $ is the transformation stress of inclusion
$M$ and $Z^K_{ij}$ is the configurational force type quantity conjugate
to $\dot{\epsilon}^{*K}_{ij}$ given by
\begin{equation}
Z^K_{ij}=\int_{V^{K}} \left ( \hat{\sigma}_{ij}+\sum_{M=1}^N
\sigma^{T_M}_{ij} \right ) dV
\label{eqzz}
\end{equation}
Direct calculations of dissipation and dissipation rate for
\cite{Eshelby57,Eshelby59} inclusions have been given by \cite{Vasoya19,
  Vasoya20}. 

Defining an average stress quantity in inclusion $K$ by
\begin{equation}
\bar{\sigma}^K_{ij}=\frac{1}{V^K} \int_{V^K} \sigma_{ij}^K dV
\label{eqzz1}
\end{equation}
Eq.~(\ref{eqzz}) can be written as
\begin{equation}
Z^K_{ij}= \left ( \bar {\hat{\sigma}}_{ij}+\sum_{M\ne K}
\bar{\sigma}^{T_M}_{ij}  + \sigma^{T_K}_{ij}  \right ) V^K
\label{eqzz2}
\end{equation}
since $\sigma^{T_K}_{ij} $ is uniform in $V^K$. Also, if the inclusion
size is sufficiently small compared to the distance over which
$\hat{\sigma}_{ij}$ and $\sigma^{T_M}_{ij}$ vary, the average values
  in Eq.~(\ref{eqzz2}) can be well-approximated by their values at
 the inclusion center.

Focusing attention on \cite{Eshelby57,Eshelby59} inclusion $K$, 
a non-negative dissipation rate is
guaranteed by a linear relation of the form
\begin{equation}
\dot{\epsilon}^{*K}_{ij}={\cal K}^{K}_{ijkl} Z_{kl}^{K}
\label{eqz1}
\end{equation}
where $Z_{kl}^{K}$ is given by Eq.~(\ref{eqzz}) and the
 tensor ${\cal K}^K_{ijkl}$ is
positive semi-definite, i.e. satisfies $B_{ij} {\cal K}^K_{ijkl} B_{kl} \ge 0$
for any second order tensor ${\bf B}$.

For an isotropic relation in
Eq.~(\ref{eqz1}), ${\cal K}^{K}_{ijkl}$ has the form 
\begin{equation}
{\cal K}^{K}_{ijkl}=\frac{1}{2} \kappa^{K}_1 \left ( \delta_{ik} \delta_{jl}+
  \delta_{il} \delta_{jk} \right ) +\kappa^{K}_2 \delta_{ij}\delta_{kl}
\label{eqz2}
\end{equation}
so that
\begin{equation}
\dot{\epsilon}^{*K}_{ij}=\kappa^{K}_1 Z^{K}_{ij} +\kappa^{K}_2 Z^{K}_{kk} \delta_{ij}
\label{eqz3}
\end{equation}
where $\delta_{ij}$ is the Kronecker delta.

The dissipation rate in inclusion $K$ can now be written as
\begin{equation}
\dot{\cal D}^K=Z^K_{ij} \dot{\epsilon}^{*K}_{ij}=\kappa^{K}_1
Z^{K}_{ij} Z^K_{ij} +\kappa^{K}_2 \left ( Z^{K}_{kk} \right )^2 \ge 0
\label{eqz5}
\end{equation}
and the total dissipation rate is
\begin{equation}
\dot{\cal D}=\sum_K \dot{\cal D}^K
\label{eqz52}
\end{equation}

 Non-negative $\dot{\cal D}$ does not require each $\dot{\cal D}^K$ to
be non-negative.  If one $\dot{\cal D}^K$ is negative, the
dissipation rate in a sub-volume including only that
inclusion is negative so that the Coleman-Noll postulate \citep{CN64}
is then not satisfied. 

\subsubsection{Plane strain shear transformation}
\label{sec:shear}

As an illustration of the general expressions, the relations for plane strain
shear of a circular inclusion are presented as used, for example, in the
numerical 
STZ calculations of \cite{Vasoya20,Vasoya21}. In the numerical
calculations, it is assumed that the STZ is sufficiently small 
relative to the length scale over which stresses vary so that the
stress state in the STZ can be regarded as uniform.

With the transformation strain in STZ $K$ being 
$\gamma^{*K} (s_in_j+s_jn_i)/2$, where $s_i$ gives the shear direction
and $n_i$ is normal to $s_i$, $Z^K_{sn}$  is given by \citep{Vasoya20}
\begin{equation}
Z^K_{sn} =   \frac{E}{2(1-\nu^2)}  \left [ \frac{ 4 {\sigma}^K_{sn}
       (1-\nu^2)}{E} - \gamma^{*K}   \right ]  A^K_{\rm incl} 
\label{eqss0}
\end{equation}
where $A^K_{\rm   incl}$ is the inclusion area.

  The dissipation rate associated with the transformation of STZ $K$ is
$ \dot{{\cal D}}^K =Z^K_{sn} \dot{\gamma}^{*K}$ and with the kinetic
relation for $\gamma^{*K}$  taken as  
\begin{equation}
\dot{\gamma}^{*K} = \frac{1}{B^K} \left [ \frac{1}{A_{\rm incl}}
  \frac{2 (1-\nu^2) Z^K_{sn}} {E} \right ] = \frac{1}{B^K}  \left [  
\gamma^{*K}_{\rm max} -  \gamma^{*K} \right ] 
\label{eqss2}
\end{equation}
where 
 
\begin{equation}
\gamma^{*K}_{\rm max}= \frac{4 {\sigma}^K_{sn}(1-\nu^2)}{E} 
\label{eqgm}
\end{equation}

 The dissipation rate is given by 
\begin{equation}
\dot{{\cal D}}^K =\frac{1}{B^K}    \left ( \frac{E A_{\rm incl}}{2(1-\nu^2)} \right ) 
\left (\gamma^{*K}_{\rm max} - \gamma^{*K} \right )^2 
\label{eqss1z}
\end{equation}
and is guaranteed to be non-negative.

From Eq.~(\ref{eqgm}),
$\gamma^{*K}_{\rm max}$ is of the order of the STZ stress divided by
an elastic modulus. Quite generally, the limiting strain magnitude for
a non-negative dissipation rate or even non-negative total dissipation
is of the order of the \cite{Eshelby57,Eshelby59}  inclusion stress
level divided by a 
relevant elastic modulus, \cite{Vasoya19}.

\begin{table}[htb!]
\begin{center}
\begin{tabular}{| c | c | c | c | c | c | c | c |}
\hline
 Composition & E (GPa) &$\nu$ & $\sigma_y $ (GPa)  & Vol. $\dot{\cal D} \ge 0$  & Shear $\dot{\cal D} \ge 0$ \\  
  \hline
Nd$_{60}$Al$_{10}$Fe$_{20}$Co$_{10}$ &51.0& 0.342& 0.45& 0.00295 & 0.0128 \\
\hline 
Ho$_{55}$Al$_{25}$Co$_{20}$ &67.0& 0.34& 0.87 &0.0043 & 0.0188 \\
\hline
Zr$_{48}$Be$_{24}$Cu$_{12}$Fe$_{8}$Nb$_8$ & 95.7& 0.359& 1.6  & 0.0054 &0.0242 \\
\hline
 Zr$_{46}$Cu$_{46}$Al$_{8}$ & 96.0& 0.37 &1.67 &  0.0055 & 0.0253\\
\hline
Fe$_{50}$Cr$_{15}$Mo$_{14}$C$_{15}$B$_6$ &217.0& 0.323& 4.17 & 0.0065  & 0.0277\\
\hline
Ti$_{45}$Zr$_{20}$Be$_{35}$ &96.8&0.356 & 1.86  & 0.0062 & 0.0279\\
\hline
Cu$_{50}$Hf$_{43}$Al$_7$ &113.0& 0.345 & 2.2 & 0.0064 & 0.0281\\
\hline
Zr$_{52.5}$Cu$_{17.9}$Ni$_{14.6}$Al$_{10}$Ti$_5$ & 88.4&0.374 & 1.91 & 0.0068 & 0.0324\\
\hline
Ca$_{65}$Li$_{9.96}$Mg$_{8.54}$Zn$_{16.5}$&23.0& 0.277 & 0.53 & 0.0083 & 0.0329 \\
\hline
\end{tabular}
\end{center}
\caption{Condensed version of the table of \cite{Vasoya19} showing the
  limiting transformation strains for a single transforming inclusion
  to have a non-negative dissipation rate 
  with a uniform remote stress. The values of $E$, $\nu$ and
  $\sigma_y$ are taken from \cite{Qu}.}
\label{tab1}
\end{table}

The data in Table~\ref{tab1} is taken from \cite{Vasoya19} and shows
the limiting transformation strains for a non-negative dissipation
rate for a shear transformation and for a volumetric transformation
with the transformation stress taken to correspond to the macroscopic
stress at the onset of plastic yielding. The limiting  transformation strain
magnitudes for a non-negative dissipation rate are all less than
$0.035$. Atomistic calculations, 
\cite{Dasgupta,Albert}, and experiments, \cite{Argon,Hufn16}, are,
when interpreted in terms of a transforming \cite{Eshelby57,Eshelby59}
inclusion, consistent with transformation strains $ \gamma^{*K}$ that
are greater than  $\gamma^{*K}_{\rm max} $. 

There are several possible reasons for the discrepancy between the
limiting strain values in Table~\ref{tab1} and the atomistic
calculations and experiment.  The limiting values of transformation
strain in Table~\ref{tab1} were calculated assuming linear elasticity
and  identifying the stress at 
transformation with the yield strength. Local stress concentrations
would increase the stress magnitude at the nucleation site and thus
lead to an increased allowable transformation strain. It could be
necessary to account for entropy, either 
temperature related entropy or configurational related entropy,   as in
a continuum STZ plasticity formulation, \cite{Bouch09,Falk11}. 

  One possibility, considered by \cite{Vasoya20}, is that the term
$\Theta \dot{s}$ in  Eq.~(\ref{cd4}) associated with a transformation
  is sufficiently positive 
for the Clausius-Duhem inequality to be satisfied even if the
mechanical dissipation is negative. Another possible 
consequence of accounting for the entropy
increase associated with  
the transformation is having  different elastic moduli in the
STZ ((through the entropy
dependence of the energy $\phi$) from those in the material outside. 
The results of  
\cite{Vasoya19}  for a  single
transforming  STZ in an infinite isotropic elastic solid with a
uniform applied 
stress $\sigma_{ij}^0$ at infinity can give a qualitative indication
of the effect.

 For a plane strain shear transformation of a
circular STZ with a mismatch in Young's modulus $E$ between the
material inside the STZ and that outside the STZ, but with the
same value of Poisson's ratio $\nu$ inside and outside the STZ,
the ratio of  mismatch/uniform values of the maximum transformation
shear strain  for 
non-negative dissipation/dissipation rate  at
the same value of $\sigma^{(0)}_{12}/G^{\rm out}$,  $R_{\rm   mism}$,
calculated from Eq.~(70)  of \cite{Vasoya19}, is  
\begin{equation}
R_{\rm mism}=\frac{\gamma^*_{\rm mism}}{\gamma^*_{\rm unif}}
=\frac{G^{\rm out}/G^{\rm in}+(3-4\nu)}{1+(3-4\nu)}
\label{ent10b}
\end{equation}
where $G=E/2(1+\nu)$ and $(\ )^{\rm in}$   and $(\ )^{\rm out}$
denote values 
inside and outside the STZ, respectively.
For $R_{\rm  mism}$ in Eq.~(\ref{ent10b}) to be greater than $1$,
$G^{\rm in} < G^{\rm out}$, so that the effect of entropy needs to be
to reduce the stiffness in the STZ. For example,
with $\nu=0.325$, $R_{\rm  mism}=2$ for $G^{\rm in}/G^{\rm
  out}=0.269$ and $R_{\rm  mism}=3$ for $G^{\rm in}/G^{\rm
  out}=0.156$. This suggests that in order to increase the
transformation strain for a non-negative dissipation rate by a factor
of $2$ to $3$, this entropy contribution would need to substantially
reduce the  stiffness inside the STZ, by a factor of $\approx 4 - 6$
in this simple estimate. 

Another possibility, focused on here, is that
 the transformation strain magnitude for some
transformations  exceeds that required for a non-negative
dissipation rate. If this occurs, 
requiring  a non-negative
dissipation rate (i.e. satisfaction of the Clausius-Duhem inequality)
for all points 
of the body and for all time is an overly restrictive
requirement.    For a limiting transformation strain $\gamma^{*K}_{\rm
  non} > \gamma^{*K}_{\rm max}$, an evolution equation
analogous to Eq.~(\ref{eqss2}) is
\begin{equation}
\dot{\gamma}^{*K} = \frac{1}{B^K}  \left [  
\gamma^{*K}_{\rm non} -  \gamma^{*K} \right ] 
\label{eqss31}
\end{equation}
with the dissipation rate, $\dot{{\cal D}}^K$, given by
\begin{equation}
\dot{{\cal D}}^K =\frac{1}{B^K}    \left ( \frac{E A_{\rm incl}}{2(1-\nu^2)} \right ) 
\left (\gamma^{*K}_{\rm max} - \gamma^{*K} \right )\left
(\gamma^{*K}_{\rm non} - \gamma^{*K} \right ) 
\label{eqss3z} 
\end{equation}
$\dot{{\cal D}}^K$,  is negative for $B^K>0$
and $\gamma^{*K}> \gamma^{*K}_{\rm max}$.

The possibility of a  negative dissipation rate can be
understood by 
focusing attention on a sub-volume containing only STZ $K$ and
noting that the rate form of conservation of energy requires
$\dot{W}=\dot{\Phi} + \dot{\cal{D}}^K$ for that sub-volume. The
transformation reduces stress levels in 
 the STZ and increases stress levels outside the
  STZ so that $\dot{\Phi}$ outside the STZ increases.
 If the total 
  $\dot{\Phi}=\dot{\Phi}_{\rm in}+\dot{\Phi}_{\rm out}$, which
 increases with increasing  transformation strain,  is 
  sufficiently large the rate form of conservation of energy can require
  $\dot{\cal{D}}^K<0$.

\subsection{Calculations of discrete STZ  plasticity dissipation}

There are few solutions of STZ plasticity boundary value problems
compared with those for discrete dislocation plasticity, particularly
solutions based on transformation strain relations that guarantee
a non-negative dissipation rate.  In their work on developing STZ
evolution relations that guarantee a non-negative dissipation rate,
\cite{Vasoya20} presented some results for the plane strain tension
problem  illustrated in Fig.~\ref{chi1} (but of course with no slip
planes). 

An initial distribution of potential STZ sites is specified with each
site having a specified critical shear strain
energy for activation taken from a Gaussian distribution. Also, each
potential STZ site is taken to be circular with the evolution of its
transformation shear strain as given in Eq.~(\ref{eqss2}).

In the calculations of \cite{Vasoya20}, $B^K=B$ for all STZs and the
direction of shear when as STZ is activated is taken to be a maximum
shear direction at the STZ center. The boundary value problem is
solved using the superposition formulation in
Section~\ref{sec:model}. An advantage of this formulation is that the
STZ size can be much smaller than the finite element size used to solve
for the $(\hat{ \ })$ fields so that there can be many STZs per finite
element. 

For a non-negative dissipation rate, the transformation strain magnitude
is limited to $\gamma^{*K}_{\rm max}$, which is typically in the range
$0.03$ to $0.04$, but plastic strains larger than $\gamma^{*K}_{\rm
  max}$ are observed in metallic glass shear bands,
e.g. \cite{Maass15}.   Obtaining larger shear strains with a
non-negative dissipation rate requires STZs to be able to
activate more than once. A time parameter $t_D$ is introduced so that
reactivation can only occur after a delay of $t_D$ from the previous
activation. Thus, even though each activation is constrained
to have a transformation strain sufficiently small 
to be associated with a non-negative dissipation rate,  repeated reactivation
permits larger transformation strains to accumulate in an STZ.
Also, in order to provide the possibility of plastic deformation
occurring between the initial potential STZ sites,  when an STZ site
activates, new potential STZ activation sites can 
nucleate in the vicinity of an activated site. 

\begin{figure}[htb!]
\begin{center}
\subfigure[]
{\resizebox{!}{65mm}{\includegraphics{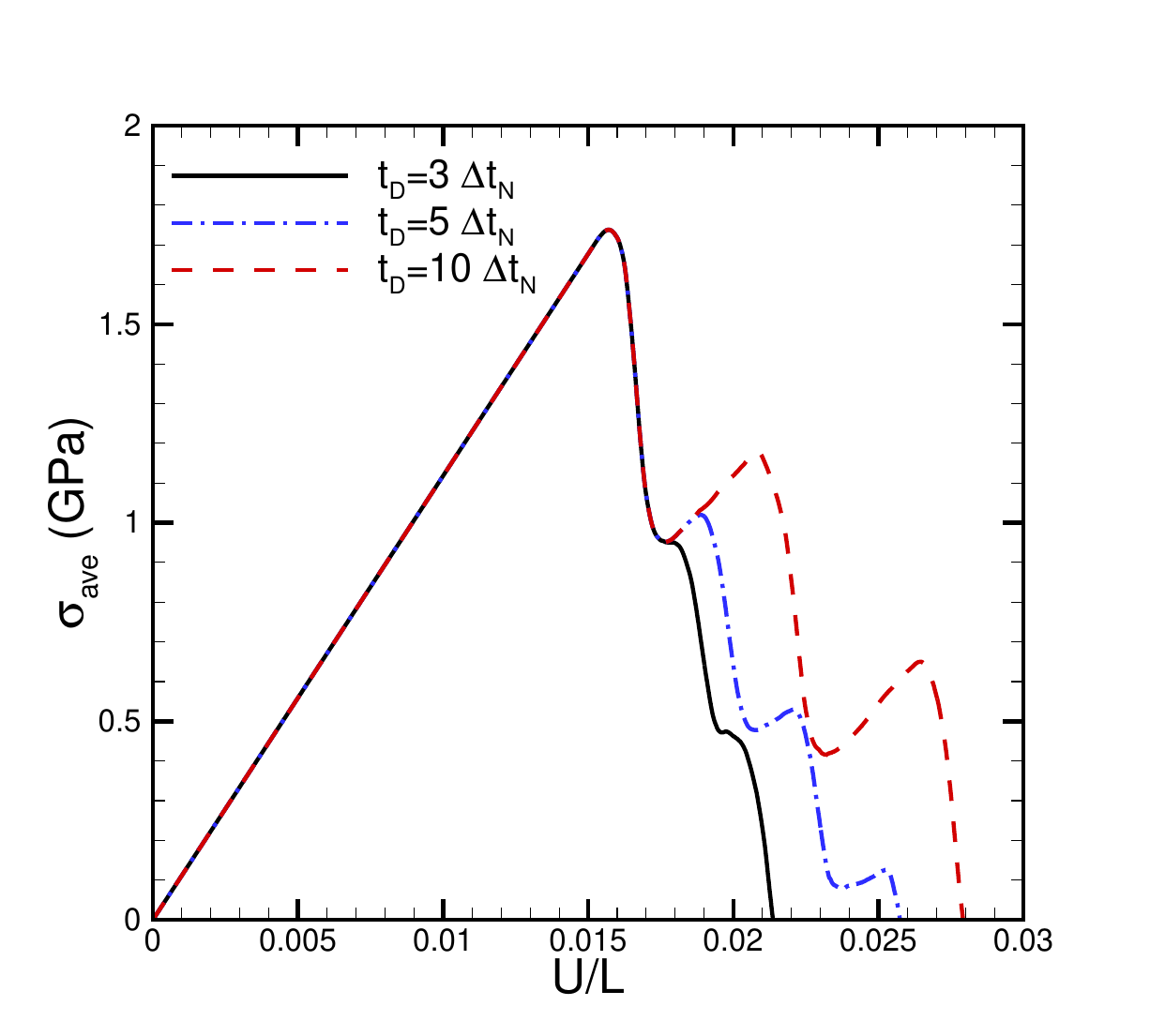}}} 
\subfigure[]
{\resizebox{!}{65mm}{\includegraphics{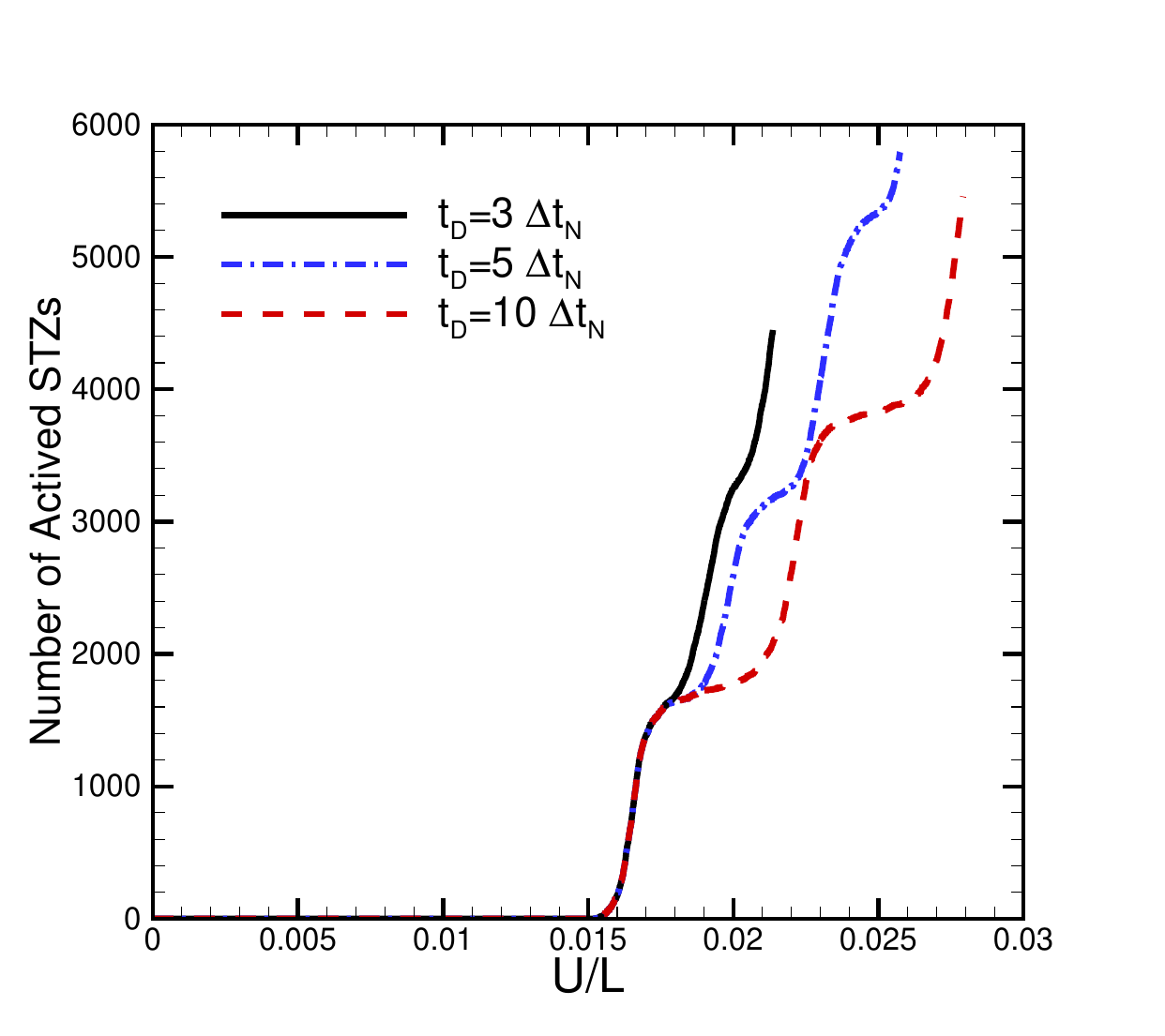}}} 
\end{center}
\caption{The dependence of the stress-strain response and the
  evolution of STZs on the delay time for reactivation $t_D$ 
  with  $B=0.001$ s. (a) Stress-strain
  curves. (b) Evolution of the number of activated STZs with overall strain,
  $U/L$. From \cite{Vasoya20}.}
\label{man20-1}
\end{figure}

\begin{figure}[htb!]
\begin{center}
{\resizebox*{120mm}{!}{\includegraphics{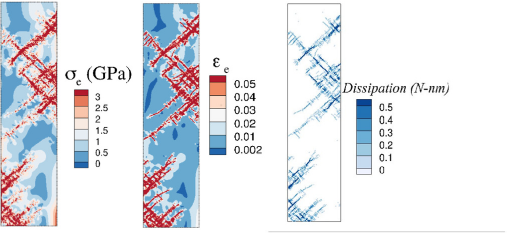}}}
\end{center}
\hspace{25mm} (a) \hspace{25mm} (b) \hspace{28mm} (c) \\
\caption{Distributions of (a) Effective stress, $\sigma_e$. (b)
  Effective strain, $\epsilon_e$. (c) Energy dissipation. Unpublished
  results based on the work of \cite{Vasoya20}.}
\label{man20-2}
\end{figure}

To illustrate the sort of responses that are obtained when implementing the
STZ kinetic relation in a numerical calculation of
STZ plasticity, results are shown for 
three values of the reactivation time $t_D$, with fixed $B$, and for
two values of $B$, with fixed reactivation time $t_D$ in
Fig.~\ref{man20-1}. 
In Fig.~\ref{man20-1}a computed stress-strain curves are shown with
$\sigma_{\rm ave}$ is the average stress on $x_2=L$ (where $U(t)$
is prescribed). As the reactivation time decreases, the
load drops more rapidly.  Fig.~\ref{man20-1}b shows the evolution of the
number of activated STZ sites.  Bursts of STZ activation occur with the
bursts associated with the drops in $\sigma_{\rm   ave}$ in
Fig.~~\ref{man20-1}a. The strain interval between bursts of 
activation increases 
with increasing $t_D$. For $t_D= 5 \Delta t_N$ and $10 \Delta t_N$
there is a clear increase in $\sigma_{\rm ave}$ between the drops,
whereas for $t_D= 3 \Delta t_N$, the activation bursts occur so close
together that the value of $\sigma_{\rm ave}$ does not increase
between them. 

Fig.~\ref{man20-2} shows distributions of effective
stress, $ \sigma_e=\sqrt{3\sigma^\prime_{ij}  \sigma^\prime_{ij}/2}$,
  where $\sigma^\prime_{ij}$ is the stress deviator
  $\sigma^\prime_{ij}=\sigma_{ij}-\sigma_{kk} \delta_{ij}/3$,
  Fig.~\ref{man20-2}a, the
  corresponding effective strain $\epsilon_e$, Fig.~\ref{man20-2}b, and plastic 
dissipation, Fig.~\ref{man20-2}c.  
Because  most of the deformation arises from the STZ transformation
strain and since there can be multiple active STZ sites 
per finite element, plotting field values on the finite element mesh
gives averages over that element which masks details of the
distributions. The distributions shown were obtained on a grid having
a mesh spacing about $1/8$ the finite element spacing which allows a
better representation of the spatial variation of the $(\tilde{\ })$ fields.
The regions with elevated effective stress, strain and dissipation 
 correspond to regions of intense STZ activity. Local values of
 $\sigma_e$ could approach $E/10$. Thus, for discrete STZ plasticity
 regions of large local dissipation do not necessarily significantly
 reduce stress  levels. 

Although, the numerical results in \cite{Kondori18,Vasoya20} show that strains
larger than the transformation strain can be reached in a shear
band by repeated reactivation, they  do not resolve
the discrepancy between the transformation strains associated with
atomistic calculations and experiment for a single transformation and
the limiting strain predicted from requiring an
\cite{Eshelby57,Eshelby59}  transformation
to have a non-negative dissipation rate.  

\section{Size dependence and dissipation in constrained shear}

\begin{figure}[htb!]
\begin{center}
{\resizebox*{45mm}{!}{\includegraphics{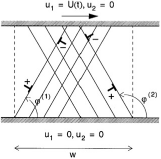}}}
\end{center}
\caption{Illustration of the discrete dislocation plasticity problem
  formulation and boundary 
  conditions for shear of a constrained layer of thickness $H$. From
  \cite{Shu01}.} 
\label{shu1}
\end{figure}

Quasi-static simple shear of a constrained strip provides a simple
context to investigate the size dependence implications of proposed
plastic material characterizations.  Length scale independent continuum 
plasticity predicts a state of uniform shear stress and uniform shear
strain that is size independent.  Discrete defect characterizations of
plastic response inevitably contain one or more length scale dependent
material parameters, such as defect size or defect spacing. However,
the material characterization containing a length scale does not
necessarily result in the overall stress-strain response being size
dependent over some range of sizes of interest. In addition, it turns
out that whether or not the 
stress-strain response exhibits size dependence can have implications
for the predicted energy dissipation. To illustrate this, 
results from \cite{Shu01} and \cite{Vasoya21} are presented. 
 Much of the focus in \cite{Shu01} and \cite{Vasoya21}  was on the size
  dependence of the stress-strain response but here the main aim in
  presenting their results is to   illustrate 
  the implications of the mode of defect evolution for the size
  dependence of dissipation. 

Fig.~\ref{shu1} illustrates the plane strain constrained shear boundary value
problem. Simple shear of a  strip, of height $H$ was analyzed with an
imposed shear displacement $U(t)$. The strip was taken to be
unbounded in the 
directions orthogonal to the height direction. The analyses were
carried out in a small deformation framework for plane strain and
quasi-static loading conditions. 

\begin{figure}[htb!]
\begin{center}
\subfigure[]
{\resizebox*{70mm}{!}{\includegraphics{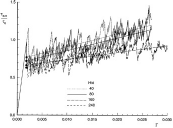}}}
\subfigure[]
{\resizebox*{70mm}{!}{\includegraphics{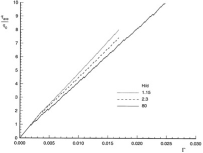}}}
\end{center}
\caption{(a) Normalized overall shear stress versus applied shear
  strain with double slip and various values of $H/d$. (b) 
Normalized overall shear stress versus applied shear
  strain with double slip and various values of $H/d$. From
  \cite{Shu01} }
\label{shu2}
\end{figure}

\begin{figure}[htb!]
\begin{center}
\subfigure[]
{\resizebox*{70mm}{!}{\includegraphics{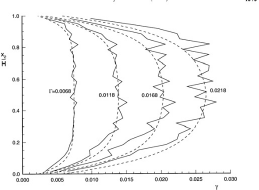}}}
\subfigure[]
{\resizebox*{70mm}{!}{\includegraphics{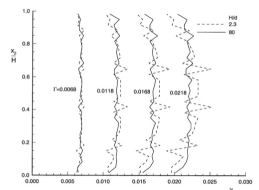}}}
\end{center}
\caption{(a) Shear strain profiles at various values of overall
  shear strain $\Gamma$ for double  slip with $H/d=80$
(b) Shear strain profiles at various values of overall
  shear strain $\Gamma$ for single slip with $H/d=2.3$ and
  $H/d=80$. From  \cite{Shu01}.}
\label{shu4}
\end{figure}

\cite{Shu01} considered planar crystals with two symmetrically
oriented slip systems as illustrated in Fig.~\ref{shu1} and cases
where dislocation sources (and therefore dislocations) were only
present on one system. The crystals analyzed had a specified
density of  initial dislocation sources with a Gaussian distribution
of  source strengths and had a specified mobility
$B^K=B$. Fig.~\ref{shu2} shows the computed overall
shear stress versus shear strain responses for the double slip,
Fig.~\ref{shu2}a,  and single slip, Fig.~\ref{shu2}b, cases. For
double slip,  there is significant size dependence and the overall slope
of the 
shear stress  versus shear strain curve in the plastic range  is much
less than the elastic 
shear modulus. It can also be seen in Fig.~\ref{shu2}a that for double
slip, presuming
nearly elastic unloading, there is significant plastic
dissipation. For single slip, Fig.~\ref{shu2}b, the slope of the
stress strain curve is little reduced from that associated with the
elastic shear modulus, very high stress levels are attained at
relatively small strains and unloading would reveal little plastic
dissipation. 

\begin{figure}[htb!]
\begin{center}
\subfigure[]
{\resizebox*{50mm}{!}{\includegraphics{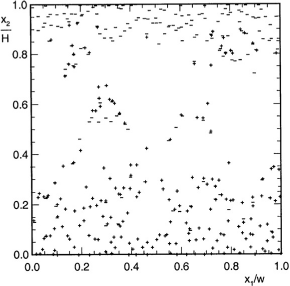}}}
\subfigure[]
{\resizebox*{50mm}{!}{\includegraphics{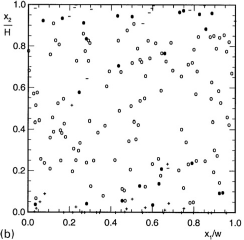}}}
\end{center}
\caption{(a)  Dislocation distributions in a unit cell with
  double slip and $H/d=80$.   (b Dislocation distributions in a unit cell with
  single slip and $H/d=80$.  From \cite{Shu01}. }
\label{shu3}
\end{figure}

Fig.~\ref{shu4}  shows strain profiles in the strip at various values
of overall shear strain, $\Gamma$, for double slip, Fig.~\ref{shu4}a,
and for single slip, Fig.~\ref{shu4}b. For double slip, there is a
clear boundary layer effect where there is little, if any, shear 
strain and the shear strain is largest in the
center of the strip. 
For single slip, the 
shear strain is nearly uniform across the strip as expected for
a material with essentially size independent  shear stress versus
shear strain response as in Fig.~\ref{shu3}b. 

The dislocation distributions at one stage of imposed shear are shown
in Fig.~\ref{shu3},  for a double slip calculation, Fig. ~\ref{shu3}a,
and for a single slip calculation, Fig.~\ref{shu3}b.  In
Fig.~\ref{shu3}a, the center of the strip, where the shear strain is
maximum in Fig.~\ref{shu4}a, is nearly dislocation free while the
edges of the strip, where there is little shear strain, have the largest
accumulation of dislocations. This is because  plastic straining
by dislocation glide is
associated with where the dislocations used to be as well as with
where  they
are. For the single slip case in Fig.~\ref{shu3}b, the distribution
of  dislocations is fairly uniform in the strip due to limited
dislocation glide, giving rise to the
rather uniform strain profiles in Fig.~\ref{shu4}b.

These results illustrate the implication for size dependence of two
possible ways to create plastic  
strain  in discrete dislocation plasticity:  (i)  nucleate dislocations 
 that glide; and (ii) nucleate dislocations that do not glide (or
 glide  very little).  

\begin{figure}[htb!]
\begin{center}
\subfigure[]
{\resizebox{!}{32mm}{\includegraphics{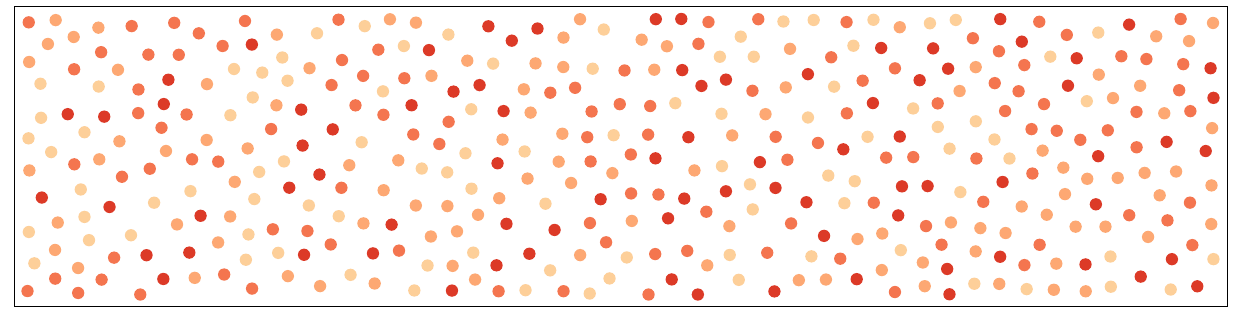}}} 
\subfigure[]
{\resizebox{!}{16mm}{\includegraphics{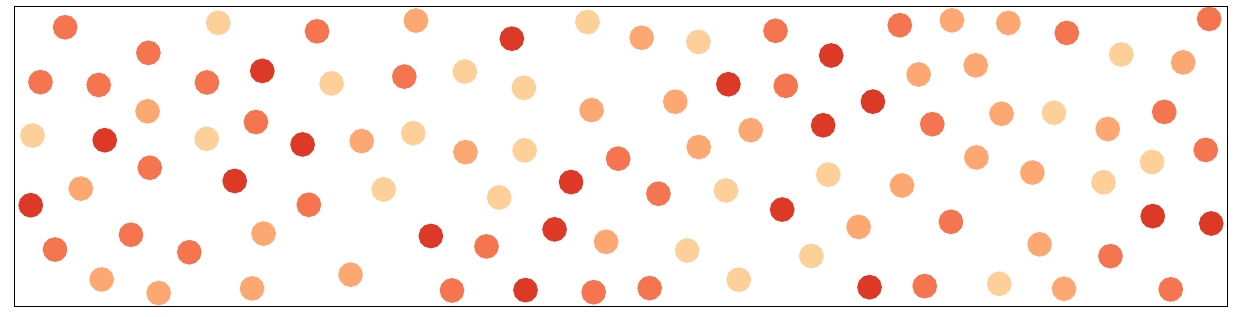}}} 
\subfigure[]
{\resizebox{!}{8mm}{\includegraphics{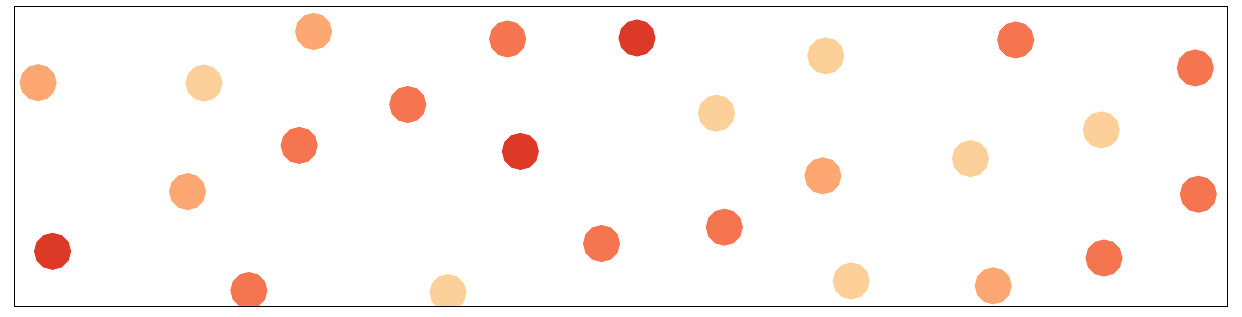}}} 
\subfigure
{\resizebox{!}{6mm}{\includegraphics{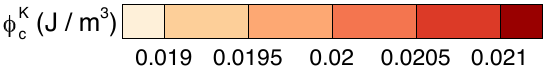}}} 
\caption{Spatial distribution of the initial potential STZ sites for
   $L/h=4$.  The shading indicates  the  value  of the critical value
  of shear strain 
energy density for STZ activation, $\phi_c^K$, associated with a given
site.  From    \cite{Vasoya21}.}
\label{mfig2}
\end{center}
\end{figure}

\cite{Vasoya21} analyzed the same constrained shear problem as
illustrated in Fig.~\ref{shu1} (but, of course, with no slip planes)
for various shear layer thicknesses (denoted by $h$ in \cite{Vasoya21})
as pictured in Fig.~\ref{mfig2}. The strip analyzed is unbounded in the
shear direction and consists of a periodic cells of length $L$. The
configurations shown in Fig.~\ref{mfig2} all have $L/h=4$.
There is an initial distribution of circular 
potential STZ activation sites. Each site is associated with a
critical value of strain energy for activation taken from a Gaussian
distribution about a  specified mean value\footnote{The value listed in
\cite{Vasoya21} was $0.02$J/m$^2$. The correct value is
$0.02$GJ/m$^2$.}. Once activated, the shear
strain evolves via Eq.~(\ref{eqss2}) so that the
dissipation rate associated with the
evolution of each STZ is non-negative. Once an
STZ activates it can nucleate additional potential STZ sites as
described by \cite{Vasoya21}. 

Fig.~\ref{mfig2} shows plots of the distribution of the
initial potential STZ sites. The shading reflects  the  value  of the
shear strain 
energy density for STZ  activation, $\phi_c^K$, associated with a
given site.  The initial locations of the STZ centers were placed
randomly. In order to exclude the possibility of partial STZs  at the  region
boundaries, 
an exclusion zone that is free of STZ centers, is introduced at each
boundary that has the width of one  STZ diameter. 

\begin{figure}[htb!]
\begin{center}
{\resizebox{!}{65mm}{\includegraphics{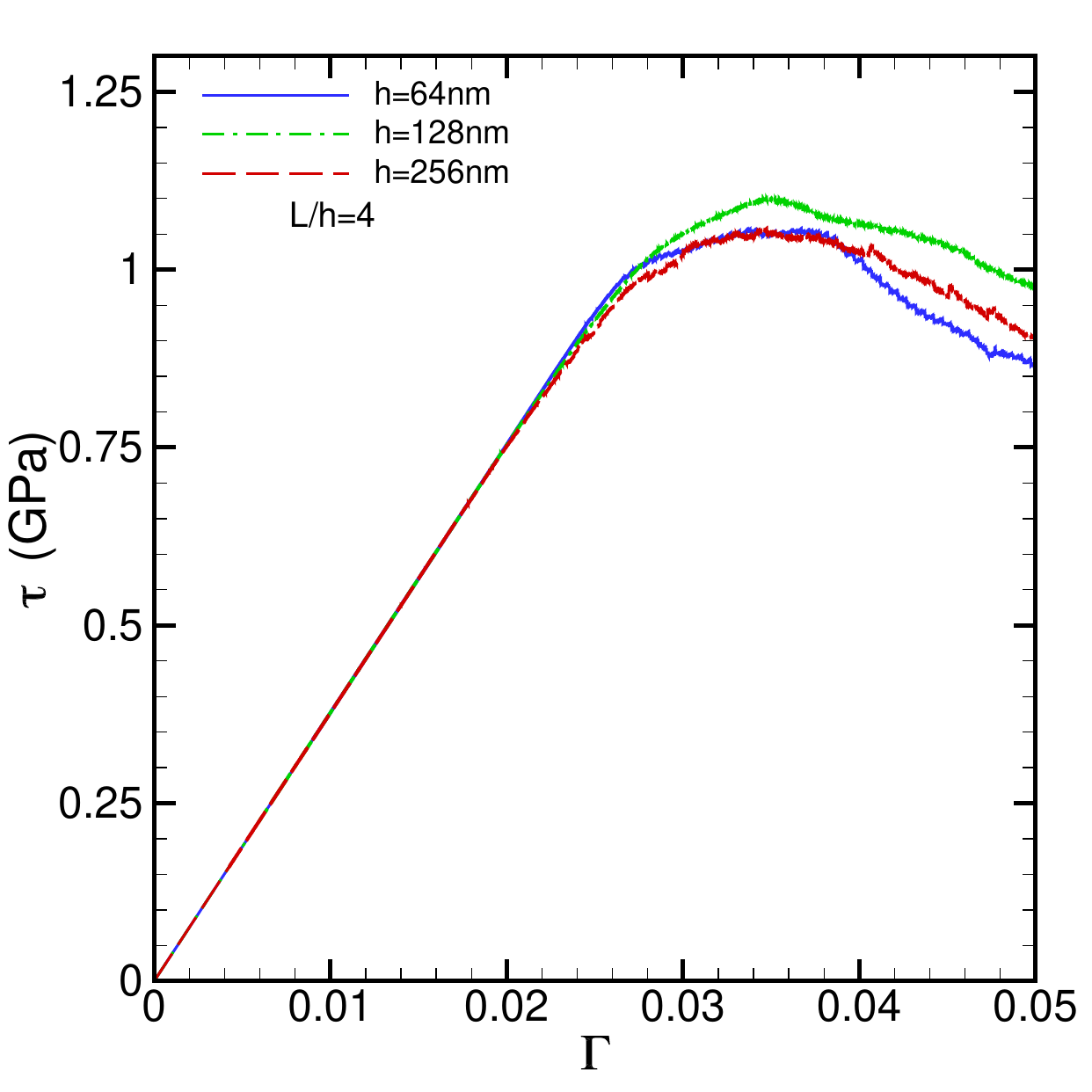}}} 
\end{center}
\caption{Evolution of shear stress $\tau$ 
and with
  overall shear strain $\Gamma$ for
   $L/h=4$ and   and for various values of $h$. From \cite{Vasoya21}.}  
\label{mfig13}
\end{figure}

\begin{figure}[htb!]
\begin{center}
\subfigure[]
{\resizebox{!}{32mm}{\includegraphics{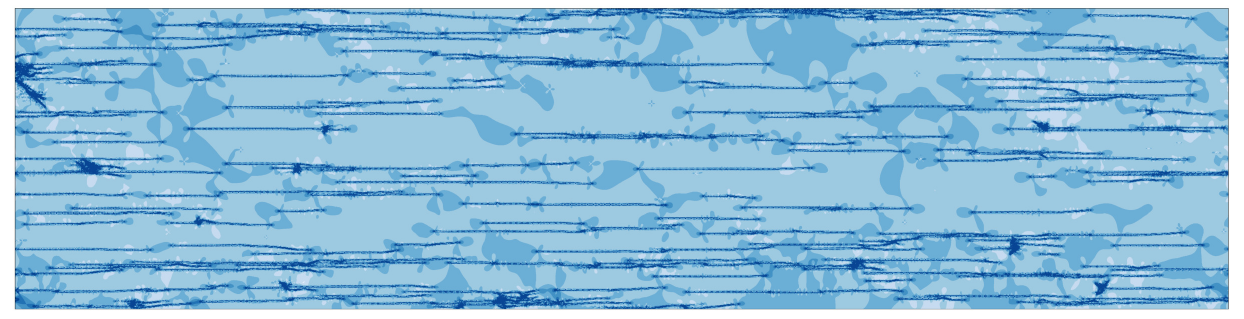}}} 
\subfigure[]
{\resizebox{!}{16mm}{\includegraphics{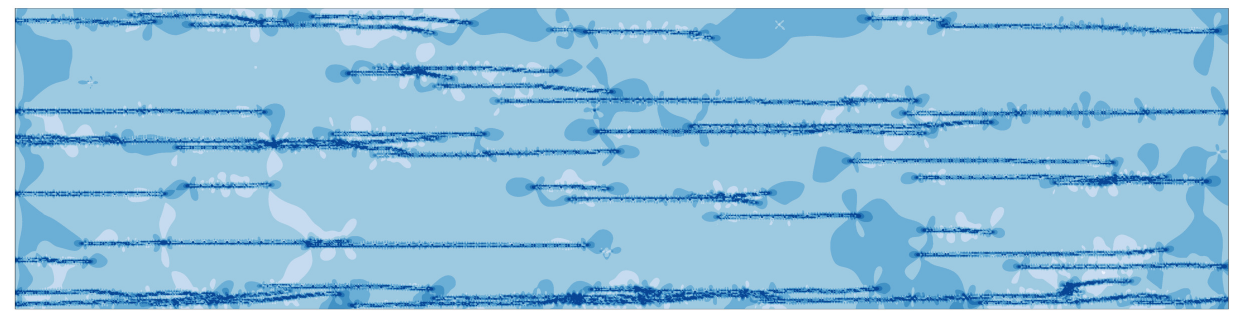}}} 
\subfigure[]
{\resizebox{!}{8mm}{\includegraphics{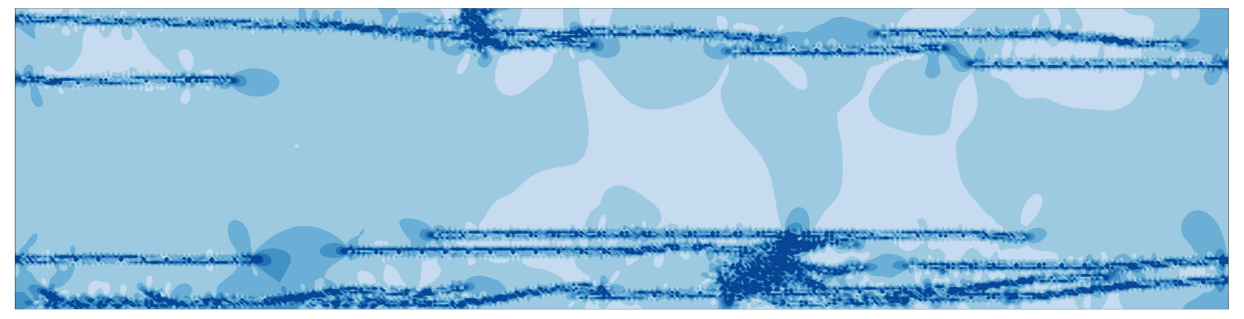}}} 
\subfigure
{\resizebox{!}{7mm}{\includegraphics{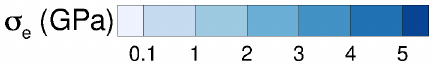}}} 
\end{center}
\caption{Distribution  of effective stress $\sigma_e$ at 
  $\Gamma=0.05$ for
$L/h=4$.
 (a) $L=256$nm.  (b) $L=128$nm. (c)   $L=64$nm. From \cite{Vasoya21}.} 
\label{mfig14}
\end{figure}

\cite{Vasoya21} analyzed two distributions of activation energies. One
was a relatively narrow distribution and the other a relatively broad 
distribution of activation energies. The results depend on the breadth
of the distribution. Here, Figs.~\ref{mfig13} and
\ref{mfig14} show some results for the relatively broad
distribution. 

Plots of the calculated overall shear stress versus
shear strain responses for several values of layer thickness $H$ are shown
in Fig.~\ref{mfig13}. There is
relatively little size dependence, within the range of what is
expected for statistical variations, and smaller is not harder. However,
the response is very different from the discrete dislocation response
response in Fig.~\ref{shu3}b where very large overall shear stress
magnitudes (relative to the initial flow strength) are reached.

The distributions of Mises effective stress in Fig.~\ref{mfig14} show
that plastic deformation takes place is well defined shear bands. The
development of shear bands is 
consistent with the softening response in Fig.~\ref{mfig13}. It is
worth noting that very large stress magnitudes, i.e. a significant
fraction of the shear modulus, occur in the shear bands. 

With the kinetic relation used by \cite{Vasoya21} which allows
for nucleation of new STZ sites, the STZ response is in some ways
intermediate between the two discrete dislocation cases of
\cite{Shu01}.  The STZs do not move through the material  as do
gliding dislocations but the 
ability to nucleate new potential STZ sites leads to a
percolation-type extension of  plastic deformation into new material
locations. This leads to a reduction in overall stress as for 
gliding dislocations but there is relatively little size dependence
as for the calculations with dislocations that do not move or move
very little.   

Although the dissipation rate was not explicitly calculated in the
results shown here, it is evident from the overall stress-strain
responses that the effect of size on dissipation is different in each
case.  Fig.~\ref{shu4}a is a case where  defects 
move relatively long distances and there is significant plastic
dissipation and a size effect, Fig.~\ref{shu4}b is a case where
defects are stationary and nucleation sites are fixed leading to
 little plastic dissipation and essentially no size effect, and
Fig.~\ref{mfig13} is a case where defects do not move but there is a
percolation-like nucleation of new defect sources leading to
significant plastic 
dissipation but no significant size effect.

\section{Continuum mechanics and a negative dissipation rate}
\label{non-pos}

Violation of the Clausius-Duhem inequality   over a small region
  and for a short time has been seen in
atomistic  simulations, for example  \cite{Ayton01}, in 
macro scale discrete particle simulations of a granular solid
\citep{Ostoja20}, and in experiment, for example  \cite{Wang10}. In
this regard it is worth noting that molecular dynamics calculations
involve an adjustable length scale parameter, typically the particle
spacing. While that spacing is regarded as an atomic spacing that is  set
to be of the order of Angstroms and all other material quantities have a length
scale that is relative to that set value. The qualitative features that
emerge from the molecular dynamics calculations would hold if the length
scale were identified with a larger length scale, of the order of
meters or kilometers. Indeed, the analyses of \cite{Ostoja20} were for
two-dimensional Couette flow of continuum scale elastic particles 
interacting via a Hookean contact relation and illustrate that
 a local  violation of the Clausius-Duhem inequality can be 
associated with discrete events regardless of absolute size. 

The extent, if any, to which  a local violation of the
Clausius-Duhem inequality is relevant for discrete defect plasticity
remains to be determined.  However, discrete defects events such as
STZ activation, phase transformation, etc. can involve relatively few
discrete entities,  take place over
relatively short times and can be influenced by fluctuations. In this
context, ``large'' and ``small'' are 
reasonably associated with the number of discrete entities involved in
a discrete defect plasticity event. What is a ``long time'' and what
is a ``short time'' are not so clear. 

 Typically, in 
formulating continuum constitutive relations, the form and parameters
of a constitutive relation are chosen to satisfy the Coleman-Noll
postulate \citep{CN64} so that the Clausius-Duhem
inequality is satisfied for all regions of a body  and for all time. In
Section~\ref{sub1d} we explore, in a simple one-dimensional
continuum mechanics context,  the consequences of violating 
the Coleman-Noll postulate \citep{CN64}, i.e. in a purely mechanical
framework having a local negative
dissipation rate over a small region and for a small time. Note that
even if there is justification in some circumstances for adding a
sufficiently positive $\dot{s} \Theta$ term so that the Clausius-Duhem
inequality is satisfied at each point, the challenge of carrying out a
stable mechanical calculation with a negative dissipation rate remains.

\subsection{One-dimensional model with a negative dissipation
  rate}
\label{sub1d}

Although the focus of the discrete
defect framework is on quasi-static deformations, the consequence of 
violating the 
Clausius-Duhem inequality (in the sense of having a negative
dissipation rate) over a small region  and for a short time   is  
explored in a one-dimensional wave propagation analysis because
dynamics provides a simple way to  introduce a length scale and a time
scale. 
The non-dimensional interval $0 \le x \le  1$ is considered with
\begin{equation}
\frac{ \partial \sigma}{\partial x}=\rho\frac{ \partial^2  u}{\partial
  t^2}
\label{w1}
\end{equation}
with the boundary conditions
\begin{equation}
\dot{u}(0)=V(t) \qquad \dot{u}(1)=0
\label{wbc1}
\end{equation}

The elastic-viscoplastic constitutive relation is
\begin{equation}
\dot{\sigma}=E(\dot{\epsilon}-\dot{\epsilon}^p)
\end{equation}
where
\begin{equation}
\dot{\epsilon}=\frac{\partial \dot{u}}{\partial x}
\label{w3}
\end{equation}
and the plastic strain rate is given by
\begin{equation}
\dot{\epsilon}^p=\dot{a} \left ( \frac{ \vert \sigma \vert}{\sigma_0}
\right )^m {\rm sign} (\sigma) 
\label{w4}
\end{equation}
The calculations are carried out using non-dimensional material
parameters with $E=1$, $\rho=1$, $\sigma_0=0.002$ and $m=0.10$. 

The imposed tensile velocity is
\begin{equation}
V(t)=\begin{cases}
 - V_0 \, t/t_{\rm rise}, &\mbox{for } t \le t_{\rm rise} 
 \\
- V_0 \,  &\mbox{for } t > t_{\rm rise} 
\end{cases}
\label{wvbc}
\end{equation}
with $V_0=1$ and $t_{\rm rise} =0.002$.

The value assigned to the parameter $a$ governs whether or not the
dissipation is positive. We specify
\begin{equation}
\dot{a} =\begin{cases}
 \dot{a}_1 \, & \mbox{for }  \vert \epsilon^p \vert \le \epsilon_c \\ 
\dot{a}_2 \, & \mbox{when} \ \vert \epsilon^p \vert = \epsilon_c  \ \mbox {and
} t \le t_0 + t_c \\
\dot{a}_1 \, & \mbox{for } t> t_0+t_c
\end{cases}
\label{abc}
\end{equation}
where $t_0$ is the time at which $\vert \epsilon^p \vert = \epsilon_c
=0.001$ and $t_c=0.15$. Also, $\dot{a}_1=1.0$ and three values of 
$\dot{a}_2$ are considered, $\dot{a}_2=0.005$, $\dot{a}_2=-0.005$ and
$\dot{a}_2=-0.01$. 

\begin{figure}[htb!]
\begin{center}
\subfigure[]
{\resizebox*{70mm}{!}{\includegraphics{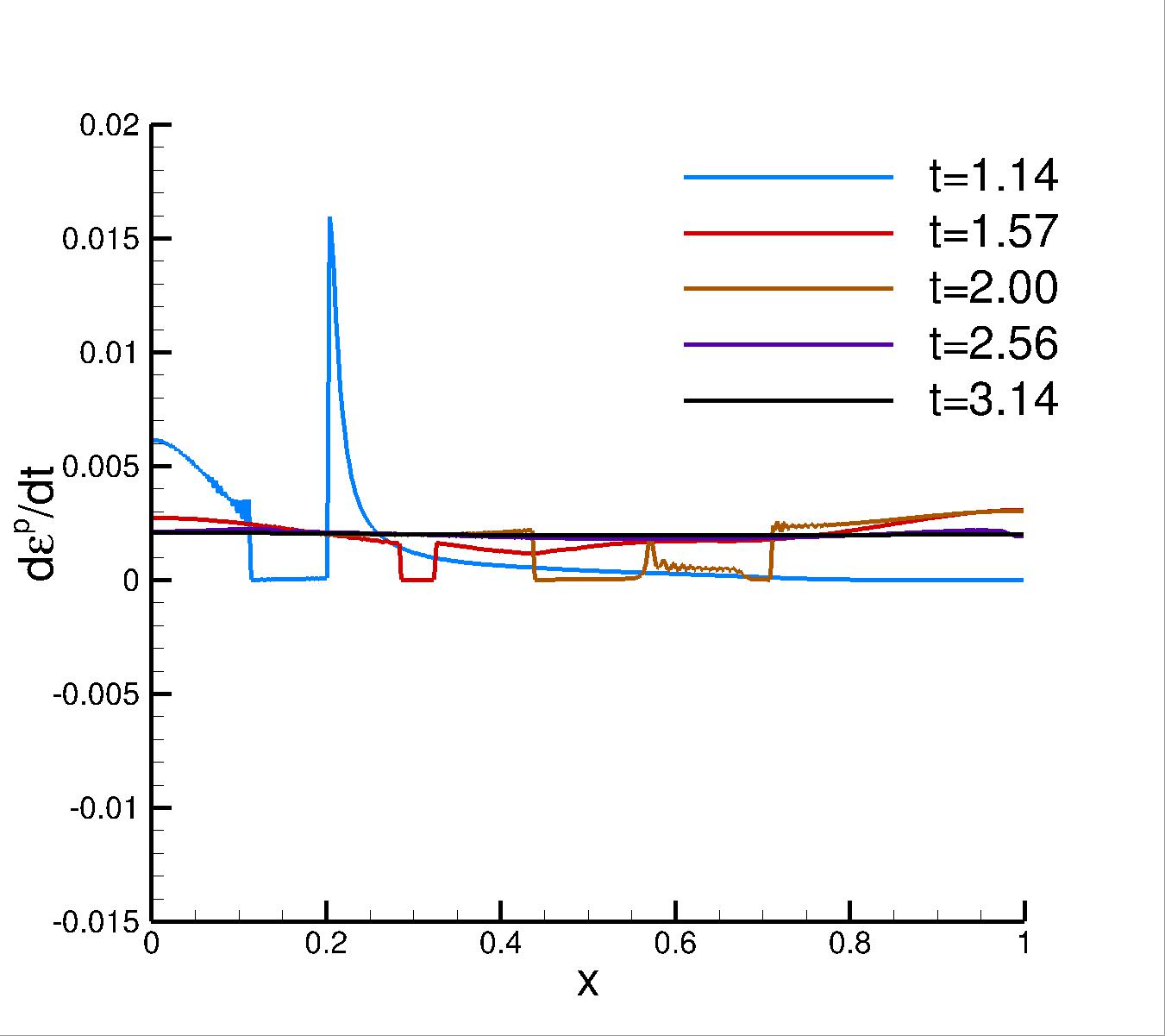}}}
\subfigure[]
{\resizebox*{70mm}{!}{\includegraphics{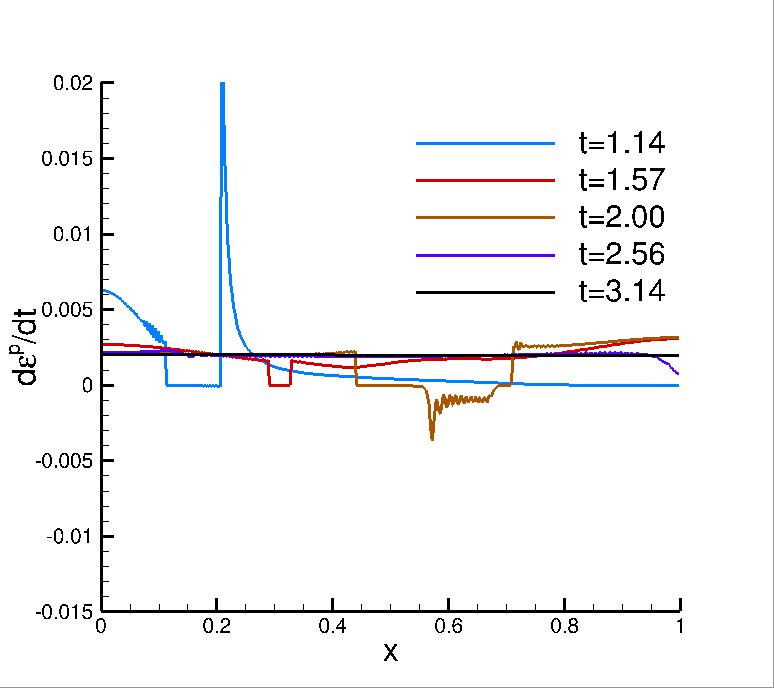}}}
\subfigure[]
{\resizebox*{70mm}{!}{\includegraphics{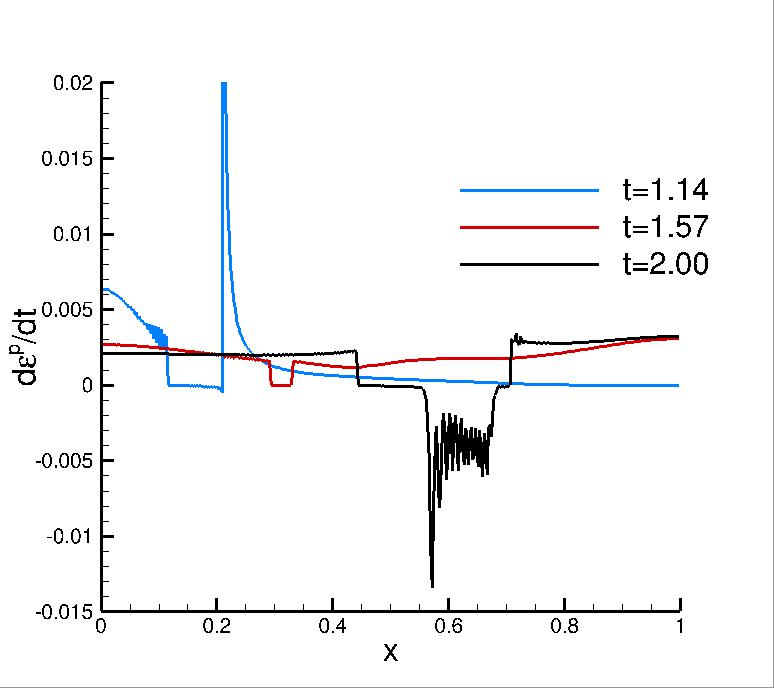}}}
\end{center}
\caption{ Plastic strain rate as a function of position at various
  normalized times. (a)   $a_2=0.005$. (b) $a_2=-0.005$. (c) $a_2= -0.01$.}
\label{1da}
\end{figure}

Time integration uses the explicit Newmark $\beta$ method with
$\beta=0$, $\gamma=0.5$ and the
stress update takes place via a linear incremental update
\begin{equation}
\sigma(t+\Delta t)=  \sigma(t) +\dot{\sigma}(t) ( \Delta t )
\label{w5}
 \end{equation}

Assuming that  plastic deformation makes no contribution to the
internal energy, the local dissipation rate, Eq.~(\ref{cd12}),  is given by  
\begin{equation}
\dot{\xi}=\vert \sigma \vert \dot{\epsilon}^p= \vert \sigma \vert
\dot{a} \left ( \frac{ \vert   \sigma \vert}{\sigma_0} \right )^m  
\label{w6}
\end{equation}
From the expression for the dissipation rate, $\dot \xi$, in
Eq.~(\ref{w6}) it is seen that the sign of $\dot \xi$ is the sign of
$\dot{\epsilon}^p$ and hence of $\dot{a}$. 
For the calculations with $\dot{a}_2<0$, there is a time interval over
which the dissipation rate is negative so that the Clausius-Duhem inequality
is violated.  

Plots of plastic strain rate $\dot{\epsilon}^p$ versus position $x$ at
various times are shown for the three calculations in
Fig.~\ref{1da}. In all three plots, the spatial distribution of
$\dot{\epsilon}^p$ is qualitatively similar, although in
Fig.~\ref{1da}a, where $\dot{a}_2>0$, the peak value of
$\dot{\epsilon}^p$ i is smaller. In Fig.~\ref{1da}a, $\dot{\xi}>0$ for
all time and all values of $x$. With increasing time, 
viscoplasticity damps out the propagating waves and a quasi-static
uniform distribution of plastic strain is reached.

With $\dot{a}_2=-0.005$, Fig.~\ref{1da}b, the distribution of
$\dot{\epsilon}^p$ shows there is an interval of $x$, the size of
which is set by the propagating wave, for which 
$\dot{\xi}<0$ so that the Clausius-Duhem inequality is
violated. 
Nevertheless, with increasing time viscoplastic dissipation 
eventually dominates and a stable quasi-static distribution of
$\dot{\epsilon}^p$  develops as for the calculation with
$\dot{a}_2=0.005$  in Fig.~\ref{1da}a. On the other hand, in
Fig.~\ref{1da}c with
$\dot{a}=-0.01$, the increased negative dissipation rate for the time
interval $t_0 \ge t \ge t_0+t_c$ leads to the overall response
becoming unstable  before
$\dot{a}$ becomes positive again and viscoplastic damping can
dominate. 

Even in this simple examples whether or not a stable response will
ultimately emerge depends on a wide variety of parameters other than
$\dot{a}_2$,  such as  $t_c$ which sets the time interval over which
a negative dissipation  rate can
occur, the specification of the applied loading which affects the
interval over which a negative dissipation rate can occur, and the
material strain and strain rate hardening which govern 
the viscoplastic damping. Nevertheless, this 
simple example illustrates that a stable overall response can be
obtained 
using a conventional continuum constitutive relation that gives rise
to  a negative dissipation rate  over a small interval for a short
time.  

The effect of negative dissipation rate on propagating waves was also
investigated by \cite{Ostoja14}   through an
analysis of one-dimensional acceleration waves with the wave amplitude
governed by a Bernoulli equation evolving on a random field of
dissipation rate. While the mean of that field was positive, the
possibility of negative dissipation rate was allowed for and it was
found that the probability of a wave amplitude blow-up event varied
with the wavefront thickness, with a blow-up event being less probable
for a thicker wavefront. 

\subsection{Discrete defect plasticity with a negative dissipation
  rate} 

The strong limitation on STZ transformation strains imposed by
requiring a non-negative dissipation rate together with the results of
atomistic modeling and experiment indicating larger transformation
strains motivates the development of  a kinetic relation for  discrete
defect plasticity that can have a negative dissipation rate (i.e. 
violate the Clausius-Duhem inequality) locally.

To illustrate a possible such relation for plane strain discrete STZ
plasticity, an extension of Eq.~(\ref{eqss2}) is 
\begin{equation}
\dot{\gamma}^{*K}  =\begin{cases}
 \frac{1}{B_1^K} \left (\gamma^{*K}_{\rm max} - \gamma^{*K} \right ) \, &
 \mbox{for }  \gamma^{*K} \le \gamma^{*K}_{\rm c} \\ 
\frac{1}{B_2^K}   \left (\gamma^{*K}_{\rm non}-\gamma^{*K}
\right )  \, & \mbox{for} \  t_c \le t \le t_c + t_f \\ 
0  \, & \mbox{for } t> t_c+t_f   \ \mbox{or}
\ \gamma^{*K}>\gamma^{*K}_{\rm non}
\end{cases}
\label{neg1}
\end{equation}
Here, $B^K_\alpha>0$ ($\alpha=1,2$), $\gamma^{*K}_{\rm non}$ is a specified
transformation strain greater 
than the value of 
$\gamma^{*K}_{\rm max}$ given by Eq.~(\ref{eqgm}),  $\gamma^{*K}_{\rm c} \le
\gamma^{*K}_{\rm max}$, $t_c$ is the first 
time at which 
$\gamma^{*K}=\gamma^{*K}_{\rm c}$, $t_f$ is a specified time
interval during which a negative dissipation rate is allowed.

The dissipation rate is then given by
\begin{equation}
\dot{\cal{D}}^K(t)=\begin{cases}
 C^K_1 \left ( \gamma^{*K}_{\rm max}-\gamma^{*K} \right )^2 
 \   &  \mbox{for }  {\gamma}^{*K} \le {\gamma}^{*K}_{\rm c} 
 \\
C^K_2 \left ( \gamma^{*K}_{\rm max} -\gamma^{*K})  (\gamma^{*K}_{\rm non}-\gamma^{*K}
\right ) , & \mbox{for} \  t_c \le t \le t_c + t_f \\ 
0  \, & \mbox{for } t> t_c+t_f \ \mbox{or}
\ \gamma^{*K}>\gamma^{*K}_{\rm non}
\end{cases} 
\label{eq11}
\end{equation}
where 
 
\begin{equation}
C^K_\alpha=\frac{EA^K_{\rm incl} }{2(1-\nu^2) B^K_\alpha}  \qquad \alpha=1,2
\label{eq10}
\end{equation} 
Hence, when $\gamma^{*K}_{\rm max} < \gamma^{*K} < \gamma^{*K}_{\rm
  non}$ for some time period  in the  interval $t_c \le t \le t_c +
t_f$, $\dot{\cal{D}}^K<0$ during that time period.

A distribution of values of  $\gamma^{*K}_{\rm c}$ can be specified so that there
is some non-zero 
probability that $\gamma^{*K} >  \gamma^{*K}_{\rm c}$ in some STZs.
Similarly for
discrete dislocation plasticity, the
expression for dislocation 
velocity, such as  in  Eq.~(\ref{eqd3a})  or Eq.~(\ref{eqd3b}), 
can be taken to apply for a range of stress magnitudes and/or a range
of values of dislocation 
velocity via expressions analogous to Eq.~(\ref{eq11}) so that 
 some values of $\dot{\cal{D}}^K$ in
Eq.~(\ref{eqd1c}) could be permitted to take on negative
values with a specified probability. This could require the 
Clausius-Duhem inequality to be satisfied for the body as
a whole, Eq.~(\ref{cd12x}),  but  allow the local condition, Eq.~(\ref{cd12})
to be violated for some  sufficiently small time interval.

\cite{Li19, Mont21} have used data from a
lower level calculation together with a statistical mechanics
fluctuation dissipation relation as a basis for computing dissipative
evolution equations/dissipation potentials. It may be possible to
extend such an approach to: (i) use lower level calculations for the
development of discrete defect plasticity kinetic relations, and/or (ii)
use discrete defect plasticity calculations for the development of
continuum plasticity constitutive relations. 

\section{Concluding remarks}

The partitioning between defect elastic energy
storage and  defect dissipation can play a
major role in a wide variety of phenomena of technological
significance, including friction, fracture, fatigue, thermal softening
and the Bauschinger effect. The predicted
partitioning of energy and its size dependence are dependent on a wide
variety of factors including, for example, 
stress state, loading rate, loading mode, e.g. monotonic loading versus
cyclic loading, and size, as well, of course, as on the kinetic  relations used
to model defect evolution.   The few discrete defect
  calculations of the evolution of dissipation  and its possible size
  effect that have been carried out show that discrete defect
  plasticity can provide insight into mechanical behaviors where the
  evolution of dissipation plays a significant role. No such calculations
  have been carried out that   allow for the possibility 
  that the Coleman-Noll postulate
\citep{CN64}, i.e. that the  Clausius-Duhem inequality must be
satisfied for all points of a body for all time, does not hold.  Discrete defect
calculations of possible effects  of
local violations of the Clausius-Duhem inequality on  predictions of
friction, fracture, 
fatigue, thermal softening and the Bauschinger effect merit exploration.

The partitioning between stored elastic energy and dissipation and its
size dependence can vary with the mode 
of defect evolution.  In the circumstances considered, when
defect evolution occurred by  defects that move relatively long 
distances, the energy partitioning involved significant dissipation and the
plastic response was size dependent; when defect evolution occurred by the
nucleation of more or less stationary  defects from fixed nucleation site the
energy partitioning involved little dissipation and the plastic
response had nearly no  size independence;  and when
defect evolution occurred   by the nucleation of stationary
defects from  sites that increased in a percolation-like manner, the
energy partitioning involved significant  dissipation  and the plastic
response had  little size independence.

It is worth noting that 
statistical effects in addition to fluctuations can play an important
role in discrete defect plasticity. For example, 
in both discrete dislocation plasticity and discrete STZ plasticity,
the initial conditions and defect nucleation/evolution parameters are
specified by statistical distributions,  for 
example, a  random spatial distribution of initial nucleation sites and
a random distribution
of  nucleation strengths.  Furthermore, the 
interaction of discrete  defects 
can be chaotic, as found for discrete dislocation
plasticity by \cite{chaos}. The predicted
overall response can be sensitive to details of the  statistical
variations and even for macro size material regions statistical
variations can induce local changes in defect structure that have a
large effect.  For example, variations in the location of defects
in the vicinity of  a crack tip can 
affect the predicted crack growth resistance.  
Quantitative discrete defect plasticity characterizations of the effect of 
statistical  variations on  material response and, 
in particular, of  the effect on the evolution of energy partitioning, are
needed.  

  For discrete STZ plasticity,  requiring a non-negative
  dissipation rate for all Eshelby transformations 
for all time  imposes a strong limit 
on the allowed transformation strain magnitude. Accounting for an
 associated  entropy increase could allow for larger transformation
strains by the term $\dot{s} \Theta$ being sufficiently positive so
that the Clausius-Duhem inequality is satisfied even if the mechanical
dissipation rate is negative and/or there could be a substantial entropy induced
reduction in stiffness in the STZ.  However, an STZ transformation 
involves a relatively small number of atoms and the transformation 
takes place over a relatively short period of time. Hence, it is also
possible that requiring
satisfaction of the  Clausius-Duhem inequality for all Eshelby transformations
for all time is  overly restrictive.

Because, at least in some contexts, the  Clausius-Duhem inequality
is a stability condition,  e.g. \cite{Cole67,Dafer79}, the
question arises as to whether or not  the  Clausius-Duhem inequality
is more broadly a stability condition and, if so, whether stability
(in some appropriate 
sense) can be maintained in discrete defect calculations  if the
Clausius-Duhem inequality is 
violated  locally. In the simple one-dimensional
analysis in Section~\ref{sub1d}  overall stability was maintained
when the dissipation rate was negative (but not too negative) locally
for a short time.  
 Such calculations for discrete defect plasticity remain to be
  carried out. A suitable continuum mechanics expression of the
  second law of 
thermodynamics for discrete defect plasticity that allows the
Clausius-Duhem inequality  to be violated locally (while maintaining overall
stability) remains to be developed. One possibility
is  to require $\int \sum \dot{\cal{D}}^K dt \ge 0$ or even each $\int
\dot{\cal{D}}^K dt \ge 0$ for some specified time interval  using 
relations like Eq.~(\ref{eqd1c})  or Eq.~(\ref{eq11}).

In a wide variety of contexts, continuum mechanics is being used to
model  the evolution of inelastic deformation of  ``small'' and/or
``soft'' materials and systems. In such systems,  fluctuations can play a
role and frameworks to  allow  fluctuation effects to be accounted
for in formulating continuum constitutive relations are
being developed, \cite{Ostoja16,Li19, Mont21}. Indeed, it is
now well-appreciated that assumptions underlying continuum mechanics
formulations for modeling
``large''    material regions and systems, such as the assumption
of a  size independent 
constitutive response, need to be abandoned to model 
``small'' scale material/system response.  For discrete defect plasticity
(and perhaps more broadly for  a variety of ``small/soft'' materials
and systems)   it may 
be necessary to develop continuum mechanics constitutive formulations
that abandon the 
requirement that   the Clausius-Duhem inequality is satisfied at all
points of a body for all time.

%\vspace{-2mm}

\section*{Acknowledgments}

\noindent I am grateful to Professors John Hutchinson of Harvard University,
Martin Ostoja-Starzewski of the University of Illinois at
Urbana-Champagne, Celia Reina of the University of Pennsylvania and
Manas Upadhyay of  \'Ecole Polytechnique for
their comments on and corrections of a previous draft. 

%\vspace{-3mm}

\end{document}